\documentclass[twoside,a4paper,nocrop]{ipart_v1}

\usepackage{t1enc}
\usepackage[latin1]{inputenc}
\usepackage[english]{babel}

\usepackage{amsmath,amssymb,amscd,amsthm,graphicx,ytableau,mathdots,xcolor,makecell,enumerate}
\usepackage{tikz,tikz-3dplot}
\tikzset{>=latex} 
\usetikzlibrary{shapes,arrows}
\usepackage{yfonts}

\usepackage{bbm}
\usepackage{bm}
\usepackage{mathrsfs}
\usepackage[T1]{fontenc}
\usepackage{lmodern}
\usepackage{enumerate}

\newcommand{\ket}[1]{| #1\rangle}

\numberwithin{equation}{section}

\theoremstyle{plain}

\newcommand{\al}{\alpha}
\newcommand{\be}{\beta}
\newcommand{\de}{\delta}

\newcommand{\vep}{\varepsilon}
\newcommand{\ga}{\gamma}

\newcommand{\la}{\lambda}

\newcommand{\vp}{\varphi}

\newcommand{\De}{\Delta}

\newcommand{\La}{\Lambda}
\newcommand{\Si}{\Sigma}

\newcommand{\bde}{\boldsymbol{\delta}}

\newcommand{\bk}{\mathbf{k}}
\newcommand{\bell}{\boldsymbol{\ell}}

\newcommand{\bp}{\mathbf{p}}

\newcommand{\bx}{\mathbf{x}}
\newcommand{\by}{\mathbf{y}}

\newcommand{\tK}{\widetilde{K}}

\newcommand{\tS}{\widetilde{S}}

\newcommand{\tk}{\tilde{k}}
\newcommand{\tbk}{\widetilde{\mathbf k}}

\newcommand{\hK}{\widehat{K}}

\newcommand{\CC}{{\mathbb C}}
\newcommand{\NN}{{\mathbb N}}
\newcommand{\RR}{{\mathbb R}}
\newcommand{\ZZ}{{\mathbb Z}}

\newcommand{\cE}{{\mathcal E}}

\newcommand{\cH}{{\mathcal H}}

\newcommand{\cP}{{\mathcal P}}

\newcommand{\cS}{{\mathcal S}}
\newcommand{\cT}{{\mathcal T}}

\newcommand{\cZ}{{\mathcal Z}}
\newcommand{\fS}{{\mathfrak S}}

\newcommand\bbe{\bar\beta}

\newcommand\Hsc{H_{\mathrm{sc}}}

\newcommand\Zsc{Z_{\mathrm{sc}}}

\newcommand{\Zspin}{Z_{\mathrm{spin}}}

\newcommand{\mss}{\kern 1pt}

\renewcommand{\le}{\leqslant}
\renewcommand{\ge}{\geqslant}
\newcommand{\tends}[1]{\bbuildrel{\hbox to 2em{\rightarrowfill}}_{#1}^{}}

\newcommand{\iu}{\mathrm i}
\newcommand{\e}{\mathrm e}
\newcommand{\diff}{\mathrm{d}}

\newcommand{\su}{\mathrm{su}}
\newcommand{\gl}{\mathrm{gl}}

\newcommand{\qbinom}[3]{{#1\atopwithdelims[]#2}_{\raise 3pt\hbox{$\scriptstyle #3$}}}
\newcommand{\en}{\enspace}

\newcommand{\Int}[1]{\,\mathop{\!#1}\limits^{\lower1ex\hbox{$\scriptstyle\circ$}}{}}

\theoremstyle{definition}
\newtheorem{rem}{Remark}

\newcommand{\mathclap}[1]{\hbox to0pt{\hss$\scriptstyle #1$\hss}}

\newcommand{\bsv}{\mathbf s}
\newcommand{\bq}{\mathbf q}

\newcommand{\Hsp}{H_{\mathrm{spin}}}
\newcommand{\Haux}{H_{\mathrm{aux}}}
\def\clap#1{\hbox to 0pt{\hss#1\hss}}

\def\mathclap{\mathpalette\mathclapinternal}

\def\mathclapinternal#1#2{%
  \clap{$\mathsurround=0pt#1{#2}$}}

\begin{document}

\title[Open supersymmetric Haldane--Shastry chain and associated motifs]{The open supersymmetric
  Haldane--Shastry spin chain and its associated motifs}

\author{J. Carrasco, F. Finkel, A. Gonz\'alez-L\'opez, M.A. Rodr\'\i guez}

\begin{abstract}
  We study the open version of the $\su(m|n)$ supersymmetric Haldane--Shastry spin chain
  associated to the $BC_N$ extended root system. We first evaluate the model's partition function
  by modding out the dynamical degrees of freedom of the $\su(m|n)$ supersymmetric spin Sutherland
  model of $BC_N$ type, whose spectrum we fully determine. We then construct a generalized
  partition function depending polynomially on two sets of variables, which yields the standard
  one when evaluated at a suitable point. We show that this generalized partition function can be
  written in terms of two variants of the classical skew super Schur polynomials, which admit a
  combinatorial definition in terms of a new type of skew Young tableaux and border strips (or,
  equivalently, extended motifs). In this way we derive a remarkable description of the spectrum
  in terms of this new class of extended motifs, reminiscent of the analogous one for the closed
  Haldane--Shastry chain. We provide several concretes examples of this description, and in
  particular study in detail the $\su(1|1)$ model finding an analytic expression for its Helmholtz
  free energy in the thermodynamic limit.
  
\end{abstract}

\maketitle

\section{Introduction}\label{intro}

Recent experiments involving trapped ions and optical lattices of ultracold Rydberg atoms have
made it possible to simulate spin chains and low-di\-men\-sion\-al lattice models with long-range
interactions, leading to a renewed interest in this type of fundamental quantum
systems~\cite{KCIKD09,GL14,JLHHZ14,RGLSS14,SZFHC15,HGCK16}. The quintessential example of these
models is the spin $1/2$ chain independently introduced by Haldane~\cite{Ha88} and
Shastry~\cite{Sh88}, in which the spins are uniformly arranged on a circle and the spin-spin
interactions decay as the square of their inverse (chord) distance. The relevance of this model
for theoretical and mathematical physics cannot be understated. Indeed, its importance in
condensed matter physics is well known, as one of the simplest models whose elementary (spinon)
excitations~\cite{Ha91,Ha93} can be naturally regarded as anyons in the framework of Haldane's
fractional statistics~\cite{Ha91b}. It has also found numerous applications in such fundamental
fields as the quantum Hall effect~\cite{AI94,BK09}, the theory of long-range
magnetism~~\cite{HGCK16}, or quantum transport in mesoscopic systems~\cite{BR94,Ca95}, to name
only a few. More recently, it has been found that the ground state of the $\su(n)$ generalization
of the Haldane--Shastry (HS) chain can be expressed in terms of chiral correlators of suitable
primary fields of the $\su(n)$ Wess--Zumino--Novikov--Witten model at level 1, a result that has
been extended to similar models with long-range interactions~\cite{CS10,NCS11,BQ14,TNS14}.

From a more mathematical standpoint, two key properties set the HS chain apart from other
integrable one-dimensional models, namely its close connection with a spin dynamical model and its
Yangian symmetry even for a finite number of sites. Indeed, the HS chain can be obtained from the
spin Sutherland model~\cite{Su72,HH92} in the strong interaction limit, through a mechanism
usually known as Polychronakos's freezing trick~\cite{Po93,Po94}. In essence, as the parameter~$a$
governing the strength of the spin-spin interaction in the spin Sutherland model goes to infinity
its eigenfunctions become increasingly peaked at the coordinates of the equilibrium positions of
the Sutherland scalar potential, which coincide with the HS chain sites. Thus in this limit the
dynamical and spin degrees of freedom effectively decouple, and the latter are governed by the HS
Hamiltonian. This connection can be used to compute in closed form the partition function of the
HS chain as the $a\to\infty$ limit of the quotient of the partition functions of the spin and
scalar Sutherland dynamical models~\cite{FG05}. In fact, this non-standard method for evaluating
the partition function can be readily applied to other spin chains of HS type with
rational~\cite{Po94,BFGR08epl} or hyperbolic~\cite{FI94,BFGR10} interactions, known respectively
as the Polychronakos--Frahm (PF) and Frahm--Inozemtsev (FI) chains and related to the integrable
spin Calogero~\cite{Ca71,MP93} and Inozemtsev~\cite{In96} dynamical models. The latter method has
also been extended to the $\su(m|n)$ supersymmetric versions of the HS~\cite{Ha93,BB06} and
PF~\cite{BUW99,BB09} chains, in which each site is occupied by either an $\su(m)$ boson or an
$\su(n)$ fermion.

The second characteristic feature of the HS chain (including its supersymmetric version) is its
invariance under the Yangian quantum group $Y(\gl(n))$ (for $\su(n)$ spin) even for a finite
number of sites~\cite{HHTBP92,BGHP93}, which is in fact at the root of many of the model's most
salient properties. To begin with, a direct consequence of the Yangian symmetry is the high
degeneracy of the spectrum, a fact already noted in Haldane's original paper. On a more
quantitative level, the model's eigenstates can be classified using certain representations of the
Yangian labeled by a class of skew Young diagrams known as border strips, whose dimension
coincides with the number of their associated semistandard Young tableaux~\cite{KKN97,NT98}. As it
turns out, these border strips are in a one-to-one correspondence with sequences of the binary
digits 0 and 1, which essentially coincide with Haldane's motifs~\cite{HHTBP92}. It should be
noted, however, that this elegant description of the spectrum in terms of motifs (or border
strips) and their associated Young tableaux cannot be obtained directly from the model's partition
function. Indeed, to derive this description it is necessary to infer a generalized partition
function depending polynomially on certain auxiliary variables, which reduces to the standard one
when evaluated at a suitable point. It is then shown that this generalized partition function can
be expressed in terms of skew Schur polynomials associated to border strips. Using the
combinatorial definition of the latter polynomials, it is then immediate to assign an energy to
each border strip and to relate its degeneracy to the number of associated Young tableaux (see,
e.g., \cite{BBHS07,BBH10}). This is seen to imply that the spectrum of the supersymmetric HS chain
coincides with that of a classical vertex model with local interactions and a suitably chosen
energy function. Again, the Yangian symmetry and its consequences described above also hold for
the supersymmetric PF chain~\cite{HB00,BBH10}. Remarkably, this description of the spectrum of the
supersymmetric HS and PF chains holds with minor changes if we add to the Hamiltonian of these
models a chemical potential term~\cite{EFG12,FGLR18}. As shown in the latter references, this
makes it possible to compute the thermodynamic functions of these models and analyze their
critical behavior using the (inhomogeneous) transfer matrix method.

The spin chains of HS type discussed so far are connected to the root system of the simple Lie
algebra $A_{N-1}$, since in all of them the spin-spin interactions depend only on the difference
of the site coordinates\footnote{Note, however, that the PF and FI chains are \emph{not}
  translationally invariant, since their sites are not uniformly spaced.}. The same is true for
the corresponding spin dynamical models of Calogero--Sutherland type, whose interaction potential
is a function of the difference of the particles' coordinates. Since the pioneering work of
Olshanetsky and Perelomov~\cite{OP83}, it has been known that it is possible to obtain integrable
variants of the (scalar) Calogero--Sutherland models of $A_{N-1}$ type associated to the extended
root systems of all the classical simple Lie algebras. It is then relatively straightforward to
construct (supersymmetric) spin dynamical models associated to the non-exceptional root
systems\footnote{In fact, the spin Calogero model of $C_N$ type is equivalent to the $B_N$ one,
  while the $C_N$ spin Sutherland model is a trivial special case of its $BC_N$ counterpart.
  Scalar dynamical models associated to the exceptional root system have also been considered, but
  their interest is more limited since they involve only a fixed number of particles.} $BC_N$,
$B_N$ and $D_N$, each of which gives rise to a corresponding spin chain through the freezing trick
(see, e.g., \cite{BPS95,Ya95,FGGRZ03,EFGR05,BFGR09,BFG09,BFG11,BFG13}). Of these three types of
models the $BC_N$ ones have received the most attention, in part because they contain one or two
more free parameters than the $B_N$ and $D_N$ ones, respectively. In particular, a reduction of
the HS chain of $BC_N$ type (in which the spin reversal operators are replaced by the identity)
has recently appeared as the parent Hamiltonian of certain infinite matrix product states
constructed from the chiral correlators of primary fields of a boundary conformal field
theory~\cite{TS15,BFG16}.

On the other hand, the spin Calogero--Sutherland models of $BC_N$ type and their associated spin
chains have not been studied to the same extent as their $A_{N-1}$ counterparts. Most notably,
although the partition functions of both the PF~\cite{BFGR09} and HS~\cite{EFGR05} chains of
$BC_N$ type have been computed in closed form (the latter only in the non-supersymmetric case),
till very recently a description of their spectrum in terms of suitable motifs has been
conspicuously lacking. For the PF chain such a description has just been provided in
Ref.~\cite{BS20}, building on previous work on the generalized partition function of this
model~\cite{BD17}. More precisely, each $A_{N-1}$-type motif splits into up to $N+1$ ``branched
motifs'' with different energies, whose degeneracies can be obtained through a combinatorial
formula.

The aim of this paper is to derive a complete description of the spectrum of the supersymmetric
Haldane--Shastry chain of $BC_N$ type in terms of suitable motifs. This model can be regarded as
an open version of the original (closed) HS chain, since its sites lie on the upper unit
half-circle and each spin interacts with the remaining ones and with their reflections with
respect to the circle's horizontal diameter. Our approach significantly differs from that of
Refs.~\cite{BD17,BS20}, since the structure of the partition functions of the PF and HS chains is
considerably different. In particular, while the generalized partition function of the PF chain is
a straightforward generalization of a Rogers--Szeg\H{o} multivariate polynomial, this is not the
case for the HS chain. Our starting point is instead a different ansatz for the generalized
partition function of the supersymmetric HS chain of $BC_N$ type, which reduces to the standard
one when evaluated at a suitable point. This generalized partition function is then expressed in
terms of two different variants of the classical super Schur polynomials. Remarkably, it can be
shown that each of these polynomials can be associated to an extended border strip of length $N+1$
(or, equivalently, motif of length $N$), where $N$ is the number of sites, and its energy
expressed in terms of the model's dispersion relation in the usual way. The crucial difference
with the $A_{N-1}$ case is that the allowed skew Young tableaux for these extended border strips
must have their last box filled by a fixed integer depending on the number of fermionic and
bosonic degrees of freedom. In this way we obtain a simple description of the spectrum in terms of
extended motifs and restricted Young tableaux, with a combinatorial expression for the degeneracy
of the corresponding multiplets.

The above result has important consequences in connection with some of the model's fundamental
properties, as we shall now discuss. To begin with, the existence of a motif-based description of
the spectrum strongly suggests that the twisted Yangian symmetry possessed by the
non-supersymmetric open HS chain\footnote{Although in Ref.~\cite{BPS95} only three particular
  instances of the HS chain of $BC_N$ type with uniformly spaced sites were discussed, the
  argument presented in this reference actually applies to the general case.}~\cite{BPS95} is also
present in its supersymmetric extension studied here. Another consequence of such a description,
together with the simplicity of the model's dispersion relation~\cite{FG15}, is the huge
degeneracy of the spectrum, which we have numerically checked for a relatively large number of
particles taking advantage of our simple characterization of the spectrum. We have also applied
this characterization to find a simple formula for the partition function of the $\su(1|1)$ model
for an arbitrary number of spins, from which we have derived a closed-form expression for its free
energy per site in the thermodynamic limit. For the general $\su(m|n)$ chain, our motif-based
description of the spectrum can be regarded as the first step towards determining the model's
thermodynamics via the inhomogeneous transfer matrix method successfully applied to its $A_{N-1}$
counterpart~\cite{FGLR18}.

This paper is organized as follows. In Section~\ref{HSsection} we introduce the model and outline
the computation of its partition function applying Polychronakos's freezing trick. This
computation is carried out in detail in Section~\ref{sec.PF}, after determining the spectrum of
the $\su(m|n)$ spin Sutherland model. Section~\ref{sec.Schurrev} is devoted to a brief review of the
definition of the classical skew super Schur polynomials and their connections with border strips
and skew Young tableaux. In Section~\ref{sec.genPF} we construct a generalized partition function
for the model, which is then applied in the following section to deduce a complete description of
the spectrum in terms of extended border strips and restricted supersymmetric Young tableaux. We
provide some specific examples of this general result in Section~\ref{sec.exa}, where we also
study in detail the $\su(1|1)$ model and its thermodynamics. Finally, in Section~\ref{sec.conc} we
present our conclusions and point out several avenues for further research suggested by our
results.

\section{The model}\label{HSsection}
The open ($BC_N$-type) supersymmetric Haldane--Shastry spin chain describes an array of $N$
particles, which can be either bosons or fermions, lying on the upper unit half-circle at fixed
angles~$2\theta_i\in(0,\pi)$ determined by the $N$ roots $\theta_i$ of the equation
\begin{equation}\label{roots}
  P^{(\beta-1,\beta'-1)}_N(\cos2\theta)=0.
\end{equation}
Here $\be$ and $\be'$ are two positive parameters, and $P^{(\beta-1,\beta'-1)}_N$ is a Jacobi
polynomial of degree $N$. Note that the chain sites $\e^{2\iu\,\theta_j}$ (with $j=1,\dots,N$) are
\emph{not} uniformly spaced unless the pair $(\be,\be')$ takes the values specified in
Table~\ref{tab.sites}.
\begin{table}[t]
  \centering
  \small
  \begin{tabular}[c]{|c|c|}
    \hline
    $(\be,\be')$& $\theta_j$\\
    \hline\hline
    $(1/2,1/2)$& $\pi (j-1/2)/2N$\\
    \hline
    $(3/2,1/2)$& $\pi j/(2N+1)$\\
    \hline
    $(3/2,3/2)$& $\pi j/(2N+2)$\\
    \hline
  \end{tabular}
  \medskip
  \caption{Values of the parameters $(\be,\be')$ for which the points $\e^{\iu\theta_j}$ with
    $j=1,\dots,N$ determined by Eq.~\eqref{roots} are uniformly spaced (the corresponding values
    of $\theta_j$ are listed in the second column).}
  \label{tab.sites}
\end{table}
  If $m$ and $n$ respectively denote the number of bosonic and fermionic
internal degrees of freedom, the Hilbert space of the system is the linear space
$\cS^{(m|n)}=\otimes_{i=1}^N\cS_i^{(m|n)}$ with $\cS_i^{(m|n)}=\CC^{m+n}$ spanned by the basis
vectors
\begin{equation}\label{basis}
  |s_1\cdots s_N\rangle:=|s_1\rangle\otimes\cdots\otimes|s_N\rangle,\qquad 1\le s_i\le m+n.
\end{equation}
In order to define the bosonic and fermionic degrees of freedom in $\CC^{m+n}$, consider two
complementary subsets $B,F\subset\{1,\ldots,m+n\}$ with $B=\{b_1,\ldots,b_m\}$ and
$F=\{f_1,\ldots,f_n\}$, where $b_1<b_2<\cdots<b_m$ and $f_1<f_2<\cdots<f_n\,$. In what follows we
shall accordingly call the single particle state~$\ket{s_i}$ bosonic if $s_i\in B$ or fermionic if
$s_i\in F$.

Setting $\theta^{\pm}_{ij}:=\theta_i\pm\theta_j$, the model's Hamiltonian can be taken
as\footnote{For the sake of simplicity, we shall omit in what follows the explicit dependence of
  $H$, $S_{ij}$ and $S_i$ on $m,n$ and $\vep,\vep'$.}
\begin{equation}\label{Hchain}
  H=\frac18\sum_{i\neq j}\bigg(\frac{1-S_{ij}}{\sin^{2} \theta_{ij}^{-} }
  +\frac{1-\widetilde{S}_{ij}}{\sin^{2} \theta_{ij}^{+} }\bigg)
  +\frac{1}{8}\sum_i\left(\frac{\beta}{\sin^{2} \theta_i} + \frac{\beta'}{\cos^2
      \theta_i}\right)\big(1-S_i \big),
\end{equation}
where the Latin indices (as in the sequel, unless otherwise stated) run from $1$ to $N$ and we
have set
\begin{subequations}\label{spin-op}
  \begin{equation}
    \widetilde{S}_{ij}:=S_iS_jS_{ij}\,.
  \end{equation}
  The Hamiltonian~\eqref{Hchain} depends on two types of operators implementing the long-range
  interaction among the spins. More precisely, the supersymmetric spin permutation operators
  $S_{ij}=S_{ji}$ are defined by
  \begin{equation}\label{spin-perm}
    S_{ij}|\cdots s_i \cdots s_j\cdots\rangle:=(-1)^{\nu(s_i,\dots,s_j)}|\cdots s_j\cdots s_i\cdots\rangle\,,
  \end{equation}
  where $\nu(s_i,\dots,s_j)$ is $0$ (respectively $1$) if $s_i,s_j\in B$ (respectively
  $s_i,s_j\in F$), and is otherwise equal to the number of fermionic spins $s_k$ with
  $i+1\le k\le j-1$. Likewise, the spin reversal operators $S_i$ are defined by
  \begin{equation}
    \label{spin-rev}
    S_i\ket{\cdots s_i\cdots}:=\la_{\vep\vep'}(s_i)\ket{\cdots \imath(s_i)\cdots}\,,
  \end{equation}
\end{subequations}
where $\vep,\vep'=\pm$ are two fixed signs and $\la_{\vep\vep'}(s_i)$ is $\vep$ for bosons (i.e,
for $s_i\in B$) and $\vep'$ for fermions (i.e., $s_i\in F$). Here $\imath$ is in general any
nontrivial involution leaving invariant the bosonic and fermionic sectors, i.e.,
$\imath^2=I\ne\imath$, $\imath(B)=B$ and $\imath(F)=F$. Assuming that $\imath$ has at most one
fixed point in each sector, we shall fix its action by setting
\[
  \imath(b_\al):=b_{m+1-\al}\,,\qquad \imath(f_{\be}):=f_{n+1-\be}\,,
\]
where, as in the sequel, the Greek indices are assumed to label the elements of the sets $B$ and
$F$ so that they run from $1$ to $m$ for bosons and from $1$ to $m+n$ for fermions unless
otherwise stated. The existence of fixed points of the involution~$\imath$ obviously depends on
the parity of the integers $m$ and $n$. Indeed, there is a bosonic (respectively fermionic) fixed
point if and only if $m$ is odd (resp. $n$ is odd). One can intuitively think of $\imath$ as
reversing the spin of a site, by simply relabeling the bosonic degrees of freedom according to
$b_\al\mapsto b_\al':=\al-(m+1)/2$ or the fermionic ones according to
$f_\be\mapsto f_{\be}':=\be-(n+1)/2$. (In other words,
$\big[\imath(b_\al)]'=b_{m+1-\al}'=m+1-\al-\frac12(m+1)=\frac12(m+1)-\al=-b_\al'$, and similarly
for fermions.)

\begin{rem}\label{rem1}
  As mentioned in the Introduction, the model~\eqref{Hchain} can be regarded as an \emph{open}
  version of the (supersymmetric) Haldane--Shastry chain. More precisely, the chain sites
  $z_j:=\e^{2\iu\theta_j}$ lie on the upper unit circle, and the spin at $z_j$ interacts not only
  with the remaining spins at $z_k$ (with $k\ne j$) but also with their reflections with respect
  to the real axes $\bar z_k$. Moreover, the strength of these interactions is equal to the
  inverse square of the distance between $z_j$ and the points $z_k$ and $\bar z_k$, respectively.
  Writing the last term in Eq.~\eqref{Hchain} as
  \[
    \frac18\sum_i\bigg(\frac{\be-\be'}{\sin^2\theta_i}+\frac{4\be'}{\sin^2(2\theta_i)}\bigg)(1-S_i)
  \]
  shows that the Hamiltonian~\eqref{Hchain} is obviously related to the $BC_N$ extended root
  system with elements $\theta_{ij}^{\pm}$, $\theta_i$ and $2\theta_i$, with $1\le i<j\le N$. Note
  also in this respect that the operators $S_{ij}$ and $S_i$ obey the
  algebraic relations
  \begin{subequations}\label{Weyl}
      \begin{gather}
        S_{ij}^2=I\,,\qquad S_{ij}S_{jk}=S_{ik}S_{ij}=S_{jk}S_{ik}\,,\qquad
        S_{ij}S_{kl}=S_{kl}S_{ij}\,,\label{Weyl1}\\
        S_i^2=I\,,\qquad S_iS_j=S_jS_i\,,\qquad S_{ij}S_k=S_kS_{ij}\,,\qquad
        S_{ij}S_j=S_iS_{ij}\,,
        \label{Weyl2}
      \end{gather}
  \end{subequations}
  where the indices $i,j,k,l$ take distinct values in the range $1,\dots,N$, and thus generate an
  algebra isomorphic to the group algebra of the $BC_N$ Weyl group. 
\end{rem}

The partition function of the chain~\eqref{Hchain} was evaluated in Ref.~\cite{EFGR05} in the
purely bosonic $(n=0)$ or purely fermionic ($m=0$) cases applying Polychronakos's freezing trick
\cite{Po93} to the spin Sutherland model of $BC_N$ type \cite{Ya95}. This method can be easily
generalized to the genuinely supersymmetric case $mn\ne0$, as we shall explain in the next
section. More precisely, the Hamiltonian of the $\su(m|n)$ spin Sutherland model is defined
by
\begin{multline}\label{Hdyn}
  \Hsp=-\De+a\sum_{i\neq
    j}\!\left(\frac{a-S_{ij}}{\sin^2x_{ij}^{-}}+\frac{a-\widetilde{S}_{ij}}{\sin^2x_{ij}^{+}}\right)\\
  +\sum_i\!\left(\frac{b(b- S_i)}{\sin^2x_i}+\frac{b'(b'- S_i)}{\cos^2x_i}\right),
\end{multline}
where $a,b,b'$ are real parameters greater than $1/2$, $x_{ij}^{\pm}:=x_i\pm x_j$,
$\De:=\sum_i\partial_{x_i}^2$, and $S_{ij}$, $S_i$ and $\tS_{ij}$ are defined by
Eqs.~\eqref{spin-op}. The particles can be regarded as distinguishable and confined to the
interval $(0,\pi/2)$ due to the inverse-square singularities at the hyperplanes $x_{ij}^\pm=k\pi$
and $x_i=k\pi/2$ with $k\in\ZZ$. We can thus take the system's configuration space as
\[
  C'=\{\bx:=(x_1,\dots,x_N)\in\RR^N:0<x_1<x_2<\cdots<x_N<\pi/2\}\,,
\]
with corresponding Hilbert space $\cH'=L^2(C')\otimes\cS^{(m|n)}$. The scalar version of the
Hamiltonian~\eqref{Hdyn} is obtained by replacing the supersymmetric spin exchange and reversal
operators by the identity, namely
\begin{equation}\label{Hsc}
  \Hsc=-\De+a(a-1)\sum_{i\neq j}\!\bigg(\frac{1}{\sin^2x_{ij}^{-}}+\frac{1}{\sin^2x_{ij}^{+}}\bigg)+\sum_i\!\bigg(\frac{b(b- 1)}{\sin^2x_i}+\frac{b'(b'-1)}{\cos^2x_i}\bigg),
\end{equation}
which acts on the Hilbert space $L^2(C')$. Note that $\Hsc$ coincides with the dynamical
Hamiltonian~\eqref{Hdyn} for the choices $(m|n)=(1|0)$ and $\vep=+1$ under the canonical
identification $L^2(C')\otimes\cS^{(1|0)}\cong L^2(C')\otimes\CC\cong L^2(C')\,$.

Setting $b=a\beta$, $b'=a\beta'$ we obviously have
\[
  \Hsp=\Hsc+8\mss a H(\bx)=-\De+a^2U(\bx)+O(a)\,,
\]
where $H(\bx)$ is obtained from the spin chain Hamiltonian~\eqref{Hchain} replacing the fixed
sites $\theta_i$ by the dynamical variables (coordinates) $x_i$ and
\[
  U(\bx)=\sum_{i\ne j}\left(\frac{1}{\sin^2x_{ij}^{-}}+\frac{1}{\sin^2x_{ij}^{+}}\right)
  +\sum_i\!\left(\frac{\be^2}{\sin^2x_i}+\frac{\be'^2}{\cos^2x_i}\right).
\]
As $a$ grows to infinity the particles tend to {\em freeze} at the coordinates of the equilibrium
of the scalar potential $U(\bx)$ on the configuration space $C'$. It can be shown that this
equilibrium is unique~\cite{CS02}, and its coordinates coincide with the chain sites $\theta_i$
\cite{OS02}. Thus in this limit the spin degrees of freedom decouple from the dynamical ones, and
are governed by the Hamiltonian $H(\theta_1,\dots,\theta_N)=H$. It follows that when $a\gg 1$ the
eigenvalues $E_{ij}$ of $\Hsp$ behave as
\[
  E_{ij}= E_{\mathrm{sc},i}+8\mss a E_j+o(a),
\]
where $E_{\mathrm{sc},i}$ and $E_j$ are any two energies of the scalar Hamiltonian~\eqref{Hsc} and
the spin chain Hamiltonian~\eqref{Hchain}, respectively. Let us respectively denote by $\Zspin$
and $Z_{\rm sc}$ the partition functions of the $BC_N$ Sutherland spin dynamical and scalar
models. The partition function $Z$ of the spin chain is then given by the exact expression
\begin{equation}\label{Z}
  Z(T)=\lim_{a\to\infty}\frac{\Zspin(8\mss a T)}{\Zsc(8\mss a T)}\,.
\end{equation}
This is, in essence, Polychronakos's freezing trick as applied to the chain~\eqref{Hchain}.

\section{Partition function}\label{sec.PF}
\subsection{Auxiliary operator}
In view of the freezing trick formula~\eqref{Z}, in order to compute the partition function of the
chain~\eqref{Hchain} we need to determine the spectra of the spin dynamical model~\eqref{Hdyn} and
its scalar counterpart~\eqref{Hsc}. To this end, we introduce the auxiliary operator
\begin{multline}\label{Haux}
  \Haux=-\De+a\sum_{i\neq j}\!\left(\frac{a-P_{ij}}{\sin^2x_{ij}^{-}}
    +\frac{a-\widetilde P_{ij}}{\sin^2x_{ij}^{+}}\right)\\
  +\sum_i\left(\frac{b(b-P_i)}{\sin^2x_i} +\frac{b'(b'-P_i)}{\cos^2x_i}\right),
\end{multline}
where $P_{ij}$, $P_i$ are defined by
\begin{subequations}\label{coor-op}
  \begin{align}
    (P_{ij}f)(\ldots,x_i,\ldots,x_j,\ldots)&=f(\ldots,x_j,\ldots,x_i,\ldots)\,,\label{coor-perm}\\
    (P_if)(\ldots,x_i,\ldots)&=f(\ldots,-x_i,\ldots)\label{coor-rev}
  \end{align}
\end{subequations}
and $\widetilde P_{ij}=P_i P_j P_{ij}$. The operators $\Haux$, $P_{ij}$, and $P_i$ are assumed to
act on the space $L^2(C)$ of square integrable functions defined on the whole open cube
$C=(-\pi/2,\pi/2)^N\allowbreak\subset\RR^N$. In particular, by contrast with $\Hsc$ the
configuration space of the latter operators is \emph{not} restricted to the ordered tuples in $C$.
We shall also tacitly identify in what follows $\Haux$ with its trivial extension $\Haux\otimes I$
to the Hilbert space $L^2(C)\otimes\cS^{(m|n)}$.

We next define total (i.e., acting simultaneously on a particle's coordinates and spin degrees of
freedom) permutation and flip operators $\Pi_{ij}$ and $\Pi_i$ as
\begin{equation}\label{Pi-op}
  \Pi_{ij}=P_{ij}\otimes S_{ij},\qquad
  \Pi_i=P_i\otimes S_i.
\end{equation}
Such operators obviously depend on $m,n$ and the signs $\vep,\vep'\,$, although we shall omit
these labels for the sake of conciseness. Note also that the operators $\{\Pi_{ij},\Pi_i\}$, as
well as their spin coordinate counterparts defined in Eqs.~\eqref{spin-op} and~\eqref{coor-op},
provide a realization of the Weyl group of $BC_N$ type. For fixed values of $m,n$ and
$\vep,\vep'$, let us denote by $\La$ the supersymmetric projector onto states totally symmetric
under the action of both $\Pi_{ij}$ and $\Pi_i$. The key observation at this point is that the
operator $\Hsp:\cH'\to\cH'$ can be shown to be unitarily equivalent to its symmetric extension
under $\Pi_{ij}$ and $\Pi_i$ to the space $\cH:=L^2(C)\otimes\cS^{(m|n)}$~\cite{EFGR05,BFG11}.
With a slight notational abuse, we shall henceforth identify both operators and thus study the
action of the spin dynamical Hamiltonian $\Hsp$ in the Hilbert space~$\La(\cH)$, instead of the
original one $\cH'=L^2(C')\otimes\cS^{(m|n)}$. The idea is of course to derive in this way the
spectrum of $\Hsp$ from that of the (essentially \emph{scalar}) auxiliary operator~\eqref{Haux}.
The spectrum of the latter operator can in turn be computed through the following standard
procedure:
\begin{enumerate}[i)]
\item Introduce a suitable (partial) order in an appropriately chosen subset of $L^2(C)$ spanning
  a dense subspace, and construct a (Schauder, i.e., non-orthonormal) basis in which the auxiliary
  operator $\Haux$ is upper triangular, and thus its eigenvalues coincide with its diagonal
  elements in this basis.

\item Take the direct product with $\cS^{(m|n)}$ and project onto $\La(\cH)$, thus obtaining a
  Schauder basis of $\La(\cH)$ in which $\Hsp$ is upper triangular, with the same diagonal
  elements and hence eigenvalues as $\Haux$.
\end{enumerate}
To better understand the last point, note that on $\La(\cH)$ we have $\Pi_{ij}=\Pi_i=I$, and thus
\[
  P_{ij}=S_{ij}\,,\qquad P_i=S_i\,.
\]
It follows that
\begin{equation}\label{HdHaux}
  \Hsp\La=\Haux\La=\La\Haux,
\end{equation}
since the operators $\La$ and $\Haux$ commute (indeed, $[P_{ij},\La]=[P_i,\La]=0$). In the next
section we shall implement the above procedure and compute the spectrum of $\Hsp$.

\subsection{Spectrum of the spin dynamical model}
As explained in the last section, we begin by constructing a Schauder basis of $L^2(C)$ in which
$\Haux$ is upper triangular. Consider, to this end, the function
\begin{equation}\label{phi}
  \phi(\bx)=\prod_{i<j}|\sin x_{ij}^+\sin x_{ij}^-|^a\prod_k|\sin
  x_k|^b|\cos x_k|^{b'},
\end{equation}
which is clearly an element of $L^2(C)$ invariant under permutations~\eqref{coor-perm} and
reversal~\eqref{coor-rev} of the coordinates, i.e., $P_{ij}\phi=P_i\phi=\phi\,$. For any integer
multiindex $\bp=(p_1,\ldots,p_N)$ with $p_i\in\ZZ$ consider the set $\{u_\bp\}$, where the
functions $u_\bp\in L^2(C)$ are defined by
\begin{equation}\label{up}
  u_\bp(\bx):=\e^{2\iu\mss \bp\cdot\bx}\phi(\bx)\,.
\end{equation}
Note that
\[
  p_i=p_j\implies P_{ij}u_\bp=u_\bp,\qquad p_i=0\implies P_iu_\bp=u_\bp.
\]
The subspace spanned by the elements $\{u_\bp\}$ is obviously dense in $L^2(C)$ (since
$\{\e^{2\iu\mss\bp\cdot\bx}\}$ is), and we can thus construct a Schauder basis out of it by
introducing an order. To do so, consider the application $\bp\mapsto\bar\bp$ defined by
\[
  \bar\bp:=(\bar p_1,\ldots,\bar p_N)=(|p_{i_1}|,\dots,|p_{i_N}|)
\]
where $(i_1,\dots,i_N)$ is a permutation of $(1,\dots,n)$ such that $\bar\bp$ is nonincreasing,
i.e., $\bar p_i\ge \bar p_{i+1}$ (and obviously nonnegative). We order the set of nonnegative
nonincreasing multiindices using the lexicographical order $\prec$, i.e., we write
$\bar\bp\prec\bar\bq$ if and only if the first nonzero difference $\bar p_i-\bar q_i$ is negative.
We then define a partial order in the set of integer multiindices $\{\bp\}$ by setting
$\bp\prec\bq$ if and only if $\bar\bp\prec\bar\bq$. This in turn induces a partial order in
$\{u_\bp\}$, namely $u_\bp\prec u_\bq$ if and only if $\bp\prec\bq$. As shown in
Ref.~\cite{EFGR05}, the auxiliary operator $\Haux$ is upper triangular in the basis obtained
ordering $\{u_\bp\}$ with any order compatible with $\prec$, with diagonal elements given by
\begin{equation}\label{eaux}
  (\Haux)_{\bp\bp}=\sum_i\big(2\bar p_i+b+b'+2a(N-i)\big)^2\,.
\end{equation}

Let us now turn to the second point of the procedure described at the end of the last section. To
begin with, let us define the spin wave functions
\[
  \ket{\bp,\bsv}:=\La(u_{\bp}\ket\bsv)=\La\bigl(\e^{2\iu\mss \bp\cdot\bx}\phi(\bx)\ket\bsv\bigr)\,,
\]
where $\bp\in\ZZ^N$ and $\ket\bsv:=\ket{s_1,\dots,s_N}$ is an element of the canonical spin
basis~\eqref{basis} of $\cS^{(m|n)}$. Since the span of the set $\{u_\bp\}$ with $\bp\in\ZZ^N$ is
dense in $L^2(C)$, the set of vectors $\{\ket{\bp,\bsv}\}$ with $\bp\in\ZZ^N$ and
$s_i\in\{1,\dots,m+n\}$ obviously spans a dense subspace of $\La(\cH)$. These vectors are however
not linearly independent, since from the identities
\[
  \Pi_{ij}\ket{\bp,\bsv}=\Pi_i\ket{\bp,\bsv}=\ket{\bp,\bsv}
\]
it follows that the state $\ket{\bp,\bsv}$ is invariant under simultaneous permutations and
reversals\footnote{By ``reversal'' of the $i$-th coordinate of $\bp$ and $\bsv$ we of course
  intend the mapping $(p_i,s_i)\mapsto(-p_i,\imath(s_i))$.} of the quantum numbers $(\bp,\bsv)$.
For this reason, in order to construct a basis from the set $\{\ket{\bp,\bsv}\}$ we can assume
without loss of generality that $p_i\in\NN\cup\{0\}$ and $p_i\ge p_{i+1}$ for all $i$. Similarly,
if $p_i=p_{i+1}$ we can obviously take (for instance) $s_i\le s_{i+1}$ for bosons and
$s_i<s_{i+1}$ for fermions. Indeed, if $s_i=s_{i+1}\in F$ we have
\[
  \ket{\bp,\bsv}=\Pi_{i,i+1}\ket{\bp,\bsv}=-\ket{\bp,\bsv}\implies \ket{\bp,\bsv}=0\,.
\]
Finally, $\ket{\bp,\bsv}=0$ when $p_i=0$ and $s_i\in B$ is a fixed point of the involution (``spin
reversal'') $\imath$ when $\vep=-1$, or $s_i\in F$ is a fixed point of $\imath$ when $\vep'=-1$.
Indeed, in the first case we have
\[
  \Pi_i\ket{\bp,\bsv}=\ket{\bp,\bsv}=\vep\ket{\bp,\bsv}=-\ket{\bp,\bsv},
\]
and similarly in the second one. With this observation in mind, we define the sets
$B_\vep\subset B$ and $F_{\vep'}\subset F$ by
\[
  B_{\vep}:=\{b_1,\ldots,b_{m_\vep}\}\,,\qquad F_{\vep'}:=\{f_1,\ldots,f_{n_{\vep'}}\}\,,
\]
with\footnote{We denote by $\pi(k)$ the parity of the integer $k$ (i.e., $0$ for even $k$ and $1$
  for odd $k$).}
\[
  m_\vep:=\frac12\big(m+\vep\pi(m)\big),\qquad n_{\vep'}:=\frac12\big(n+\vep'\pi(n)\big).
\]
It then follows from the above remarks that when $p_i=0$ we can restrict without loss of
generality the corresponding spin component $s_i$ to $B_\vep\cup F_{\vep'}$. Summarizing, we have
found the following necessary conditions\footnote{To be sure, condition (B2) below could actually
  be replaced by equivalent ones like, e.g., $s_i\le s_{i+1}$ if $s_i\in B$ and $s_i<s_{i+1}$ if
  $s_i\in F$.} on the quantum numbers $(\bp,\bsv)$ for the set $\{\ket{\bp,\bsv}\}$ to be a basis
of $\La(\cH)$.

\begin{enumerate}[(B1)]
\item The integer multiindex $p=(p_1,\ldots,p_N)$ is nonnegative and nonincreasing, i.e.,
  $p_i\in\NN\cup\{0\}$ and $p_i\ge p_{i+1}\,$ for all $i$.
\item If $p_i=p_{i+1}$ then $s_i\ge s_{i+1}$ if $s_i\in B$ and $s_i>s_{i+1}$ if $s_i\in F$.
\item If $p_i=0$ then $s_i\in B_\vep\cup F_{\vep'}$.
\end{enumerate}
It is straightforward to show that the above conditions are actually sufficient, i.e., that they
ensure the linear independence of the set $\{\ket{\bp,\bsv}\}$.

It follows from Eq.~\eqref{HdHaux} that the action of the spin dynamical Hamiltonian~$\Hsp$ is
upper triangular in any basis $\frak B$ of $\La(\cH)$ constructed from states $\{\ket{\bp,\bsv\}}$
with $(\bp,\bsv)$ satisfying the above three conditions, provided that we set
$\ket{\bp,\bsv}\prec\ket{\bp',\bsv'}$ if and only if $\bp\prec\bp'$. Indeed,
\begin{align*}
  \Hsp\ket{\bp,\bsv}&=\Hsp\La\bigl(u_\bp\ket\bsv\bigr)=\La\Haux\bigl(u_\bp\ket\bsv\bigr)
                      =\La\bigl((\Haux u_\bp)\ket\bsv\bigr)\\
                    &=\La\Biggl(\,\sum_{\bp'\preceq\bp}(\Haux)_{\bp'\bp}u_{\bp'}\ket\bsv\!\Biggr)
                      =\sum_{\bp'\preceq\bp}(\Haux)_{\bp'\bp}\ket{\bp',\bsv},
\end{align*}
where the symbol $\bp'\preceq\bp$ indicates that either $\bp'\prec\bp$ or $\bp'=\bp$. Of course,
if $\bp'\ne\bp$ the quantum numbers $(\bp',\bsv)$ need no longer satisfy conditions (B1)--(B3)
above (in particular, the state~$\ket{\bp',\bsv}$ could vanish). However, if $\ket{\bp',\bsv}\ne0$
applying suitable permutations and reversals to these quantum numbers we can always write
\[
  \ket{\bp',\bsv}=\pm\ket{\bp'',\bsv'}\,,
\]
with $(\bp'',\bsv')$ satisfying (B1)--(B3). Since the partial order $\prec$ is obviously invariant
under permutations and sign reversals we obviously have $\bp''\prec\bp$, and therefore
$\ket{\bp'',\bsv'}\prec\ket{\bp,\bsv}$. This indeed shows that $\Hsp$ is indeed upper triangular
in the basis~$\frak B$ of states~$\ket{\bp,\bsv}$ satisfying conditions~(B1)--(B3) and partially
ordered by $\prec$, with eigenvalues
\begin{equation}\label{ed}
  (\Hsp)_{\bp\bsv,\bp\bsv}=(\Haux)_{\bp\bp}=\sum_i\big(2 p_i+b+b'+2a(N-i)\big)^2=:E_{\bp}.
\end{equation}
Since $E_{\bp}$ does not depend on $\bsv$, each multiindex $\bp$ satisfying condition~(B1) gives
rise to an eigenvalue of $\Hsp$ whose intrinsic (or spin) degeneracy $d(\bp)$ is equal to the
number of spin configurations $\bsv$ satisfying conditions~(B2) and~(B3).
\begin{rem}
  A similar argument shows that the eigenvalues of the scalar Hamiltonian $\Hsc$ are also given by
  Eq.~\eqref{ed}, although in this case each of them has no spin degeneracy. Thus $\Hsp$ and
  $\Hsc$ have the same (distinct) eigenvalues, but with different degeneracies.
\end{rem}

In order to compute the spin degeneracy of the eigenvalues of $\Hsp$, let us divide the vector
$\bp$ in ``sectors'' consisting of equal entries, i.e.,
\begin{equation}\label{ks}
  \bp=(\,\underbrace{\pi_1,\dots,\pi_1}_{k_1}\,,\dots\,,\underbrace{\pi_r,\dots,\pi_r}_{k_r})\,,
\end{equation}
where $k_i$ is the number of entries with the same value $\pi_i$ and $\pi_1>\cdots >\pi_r\ge 0$ on
account of condition~(B1). Note that the number of sectors $r$ is always between $1$ and $N$, and
that $k_1+k_2+\cdots+k_r=N$, i.e., $\bk$ belongs to the set $\cP_N$ of compositions of the integer
$N$ (that is, partitions with order taken into account). The spin degeneracy $d(\bp)$ of the
eigenvalue $E_{\bp}$ depends only on the vector $\bk=(k_1,\ldots,k_r)$ (i.e., on the lengths of
the sectors in $\bp$) and on the value $\pi_r$ of the last (smallest) distinct entry of $\bp$.
Indeed, $d(\bp)$ is obviously a product whose factors are the different ways of ``filling'' the
spin components of $\bsv$ corresponding to each sector in $\bp$ in accordance to
conditions~(B2)-(B3) above. For each of the first $r-1$ sectors we have $\pi_i>0$, so that
condition~(B3) is vacuous. Hence in this case the number of fillings is simply equal to the number
of ways in which one can choose $k_i$ values among $m$ bosonic spins (which can appear more than
once) and $n$ fermionic ones (which cannot), i.e.,
\begin{equation}\label{dmnki}
  \sum_{l=0}^{k_i}\binom{m+l-1}{l}\binom{n}{k_i-l} =:d^{(m|n)}_{k_i}\,.
\end{equation}
The same is true for the last sector when $\pi_r>0$. On the other hand, if $\pi_r=0$ we must take
condition~(B3) into account, and hence the number of bosonic and fermionic values available to
fill the last sector of $\bp$ is reduced respectively to $m_\vep$ and $n_{\vep'}$. Hence the
number of fillings of the last sector is in this case given by $d^{(m_\vep|n_{\vep'})}_{k_r}$.
Thus the intrinsic degeneracy of the eigenvalue of $\Hsp$ associated with the multiindex $\bp$ is
given by
\begin{equation}\label{dp-def}
  d(\bp)=d_{\vep\vep'}^{(m|n)}(\pi_r,k_r)\prod_{i=1}^{r-1}d^{(m|n)}_{k_i},
\end{equation}
where $d^{(m|n)}_k$ is defined by Eq.~\eqref{dmnki} and
\begin{equation}\label{deg-def}
  d_{\vep\vep'}^{(m|n)}(\pi_r,k_r)=
  \begin{cases}
    d^{(m|n)}_{k_r},\quad &\pi_r>0,\\[8pt]
    d^{(m_\vep|n_{\vep'})}_{k_r},&\pi_r=0.
  \end{cases}
\end{equation}

\subsection{Computation of the partition function}
We are now ready to compute the partition function $\Zspin$ of $\Hsp$ in the large coupling
constant limit $a\to\infty$. To this end, given a multiindex $\bp$ of the form~\eqref{ks}
satisfying condition~(B1) let us denote by
\[
  K_j:=\sum_{i=1}^jk_i
\]
the partial sums of the vector $\bk$. Setting
\[
  \bar\be:=\frac12(\be+\be')=\frac{b+b'}{2a}
\]
and expanding Eq.~\eqref{ed} in powers of $a$ after a straightforward calculation we
obtain~\cite{EFGR05}
\[
  E_{\bp}=E_0+8a\sum_{j=1}^r \pi_jk_j\left(\,\bar\be+N-K_{j-1}-(k_j+1)/2\right)+O(1)\,,
\]
where
\[
  E_0=4a^2\sum_i(\bbe+N-i)^2=\frac23\,Na^2\big(2N^2+3(2\bbe-1)N+6\bbe(\bbe-1)+1\big)
\]
is the ground state energy of $\Hsp$ and $\Hsc$. Writing $q:=\e^{-1/T}$ and taking the limit
$a\to\infty$ we thus have
\[
  \lim_{a\to\infty}\!q^{-E_0/8a}\Zspin(8aT)=\sum_{\bk\in\cP_N}\,\sum_{\pi_1>\cdots>\pi_r\ge
    0}d(\bp)\,q^{\sum_{j=1}^r \!\pi_j k_j\left(\,\bar\be+N-K_{j-1}-(k_j+1)/2\right)}.
\]
The latter sum can be evaluated using the formula
\[
  \Si_l:=\sum_{\pi_1>\cdots>\pi_l>0}q^{\sum_{j=1}^l \!\pi_j
    k_j\left(\,\bar\be+N-K_{j-1}-(k_j+1)/2\right)}=\prod_{i=1}^l\frac{q^{\cE(K_i)}}{1-q^{\cE(K_i)}}\,.
\]
proved in Ref.~\cite{EFGR05}, where
\begin{equation}\label{disprel}
  \cE(j):=\frac12\,j(2\mss\bar\be+2\mss N-j-1)
\end{equation}
can be interpreted as the {\em dispersion relation} of the HS chain of $BC_N$ type~\eqref{Hchain}.
Indeed, taking Eqs.~\eqref{dp-def}-\eqref{deg-def} into account we easily obtain the following
asymptotic expression for the partition function of the $\su(m|n)$ supersymmetric spin Sutherland
mode of $BC_N$ type:
\begin{align}
  \lim_{a\to\infty}\!q^{-E_0/8a}\Zspin(8aT)
  &=\sum_{\bk\in\cP_N}\left(\prod_{i=1}^rd_{k_i}^{(m|n)}\cdot\Si_r
    +d_{k_r}^{(m_\vep|n_{\vep'})}\prod_{i=1}^{r-1}d_{k_i}^{(m|n)}\cdot\Si_{r-1}\right)\\
  &=\sum_{\bk\in\cP_N}\left(\,\frac{d_{k_r}^{\mss(m|n)}\,q^{\cE(N)}}{1-q^{\cE(N)}}+
    d_{k_r}^{\mss(m_\vep|n_{\vep'})}\right)\prod_{i=1}^{r-1}\frac{d_{k_i}^{\mss(m|n)}\,
    q^{\cE(K_i)}}{1-q^{\cE(K_i)}}\,.\notag
\end{align}
The partition function of the scalar Sutherland model of $BC_N$ type was computed in
Ref.~\cite{EFGR05} in the same fashion (or is just obtained from the previous expression setting
$m=1$, $n=0$ and $\vep=+1$), with the result
\begin{equation}\label{Zsck}
  \lim_{a\to\infty}\!q^{-E_0/8a}\Zsc(8aT)=\prod_{i=1}^N\Big(1-q^{\cE(i)}\Big)^{-1}.
\end{equation}
The partition function $Z(T)$ of the spin chain~\eqref{Hchain} follows from the freezing trick
formula~\eqref{Z}, namely
\begin{equation}\label{Zk}
  Z(T)=\sum_{\bk\in\cP_N}F(q,\bk)\left(d^{\mss(m_\vep|n_{\vep'})}_{k_r}
    +\left(d^{(m|n)}_{k_r}-d^{(m_\vep|n_{\vep'})}_{k_r}\right)q^{\cE(N)}\right)
  \prod_{i=1}^{r-1}d_{k_i}^{(m|n)},
\end{equation}
where $\{K'_1,\ldots,K'_{N-r}\}$ is the complement of the set $\{K_1,\ldots,K_{r}\}$ in
$\{1,\ldots,N\}$ (with $K'_1<K'_2<\cdots<K'_{N-r}$) and
\begin{equation}\label{Fq}
  F(q,\bk):=\prod_{i=1}^{r-1}q^{\cE(K_i)}\prod_{j=1}^{N-r}(1-q^{\cE(K'_j)})\,.
\end{equation}

\section{Skew super Schur polynomials}\label{sec.Schurrev}
\subsection{Symmetric polynomials}
We shall start by briefly reviewing some well-known properties of symmetric polynomials to fix the
notation (see, e.g., Ref.~\cite{Ma95} for an in-depth treatment). The complete (homogeneous)
symmetric polynomial $h_k(\bx)$ of degree $k$ in the vector variable $\bx:=(x_1,\dots,x_m)$ is
defined by
\[
  h_k(\bx)=\sum_{1\le j_1\le\cdots\le j_k\le m}x_{j_1}\cdots x_{j_k}\,.
\]
Likewise, the elementary symmetric polynomial $e_k(\by)$ of degree $k$ in the vector variable
$\by:=(y_1,\dots,y_n)$ is given by
\[
  e_k(\by)=\sum_{1\le j_1<\cdots<j_k\le n}y_{j_1}\cdots y_{j_k}\,.
\]
The generating functions for these polynomials are respectively
\begin{equation}
  \prod_{i=1}^m\frac{1}{1- tx_i} =\sum_{k=0}^{\infty}h_{k}(\bx)t^k,\qquad
  \prod_{i=1}^n(1+ty_i) =\sum_{k=0}^{n}e_k(\by)t^k.
\end{equation}
From these families of symmetric polynomials we construct the polynomials (supersymmetric
elementary functions) $E_k^{(m|n)}$ in the vector variables $\bx=(x_1,\ldots,x_m)$,
$\by=(y_1,\ldots,y_n)$ as\footnote{It is understood that $e_k(\by)=0$ for $k>n$.}
\begin{equation}
  E^{(m|n)}_k(\bx,\by)=\sum_{j=0}^{k}h_j(\bx)e_{k-j}(\by).
\end{equation}
The generating function of these polynomials is obviously
$\prod_{i=1}^m(1-tx_i)^{-1}\cdot\prod_{j=1}^n(1+ty_j)$. It is immediate to check
that\footnote{Indeed, $(1+ t)^n=\sum_{l=0}^{n}\binom{n}{l}t^l$,\enspace
  $(1- t)^{-m}=\sum_{l=0}^{\infty}\binom{m+l-1}{l}t^l$.} that the value of
$E^{(m|n)}_{k}(\bx,\by)$ at the point $\bx=(1,\ldots,1)=:(1^m)$, $\by=(1^n)$ is given by
\begin{equation}\label{Ed}
  E^{(m|n)}_{k} (1^m,1^n)=\sum_{j=0}^{k}\binom{m+j-1}{j}\binom{n}{k-j}=d^{(m|n)}_k,
\end{equation}
where it is understood that the combinatorial number $\binom{r}{s}$ vanishes for $s>r$.

\subsection{Schur polynomials}\label{sec.Schur}

\begin{figure}[t]
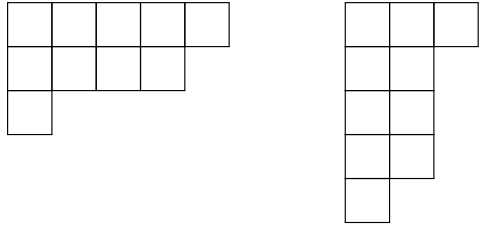

  \ydiagram{5,4,1}\qquad\qquad \ydiagram{3,2,2,2,1}
  \caption{Young diagram of shape $\la=(5,4,1)$ (left) and its conjugate $\la'=(3,2,2,2,1)$
    (right).}
  \label{fig.Ydiag}
\end{figure}

We next define the standard Schur polynomials. To this end, consider the Young diagram labeled by
an integer multiindex $\la=(\la_1,\dots,\la_r)$ with $\la_1\ge\cdots \ge \la_r>0$, which by
definition consists of $\la_1$ boxes in the first (top) row, $\la_2$ boxes in the second row,
etc.~(cf.~Fig.~\ref{fig.Ydiag}). A semistandard Young tableau of shape $\la$ is any filling of the
Young diagram $\la$ with natural numbers whose entries weakly increase along each row (from left
to right) and strictly increase down each column. The Schur polynomial $S_\la(x_1,\dots,x_m)$
corresponding to the Young diagram $\la$ is then defined by
\begin{equation}\label{skdef}
  S_\la(x_1,\dots,x_m)=\sum_Tx_1^{t_1}\cdots x_m^{\vphantom{t_1}t_m},
\end{equation}
where $T$ is any semistandard Young tableau of shape $\la$ filled with the integers
$\{1,\dots,m\}$ and $t_i$ is the number of times the integer $i$ appears in $T$. In particular,
note that $e_k=S_{(1^k)}$ and $h_k=S_{(k)}$. The polynomial $S_\la$ can be expressed in terms of
either the complete or the symmetric homogeneous polynomials through the Jacobi--Trudi
determinantal formulas
\begin{align}\label{skeh}
  S_\la(x_1,\dots,x_m)&=\det\bigl(h_{\la_i-i+j}(x_1,\dots,x_m)\bigr)_{i,j=1}^r\\
                      &=
                        \det\bigl(e_{\la'_i-i+j}(x_1,\dots,x_m)\bigr)_{i,j=1}^s,\notag
\end{align}
where $\la'=(\la'_1,\dots,\la'_s)$ is the Young diagram conjugate to $\la$ (obtained exchanging
the rows and columns of $\la$, or equivalently reflecting $\la$ about its main diagonal;
cf.~Fig.~\ref{fig.Ydiag}).

\begin{figure}[t]
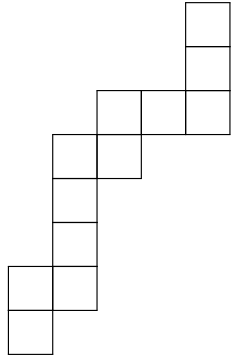

  \ydiagram{4+1,4+1,2+3,1+2,1+1,1+1,2,1}
  \caption{The border strip $\langle3,1,2,4,2\rangle=(5,5,5,3,2,2,2,1)/(4,4,2,1,1,1)$.}
  \label{fig.border}
\end{figure}

More generally, a Schur polynomial can be associated to any skew Young diagram, which we define
next. If $\lambda=(\la_1,\dots,\la_r)$ and $\mu=(\mu_1,\dots,\mu_s)$ are two Young diagrams such
that $\mu\subset\lambda$ (i.e., $s\le r$ and $\mu_i\le\la_i$ for all $i$), we define the skew
diagram $\lambda/\mu$ as the set-theoretic difference $\lambda-\mu$, obtained by removing $\mu_i$
boxes from the $i$-th row of $\la$ starting from the left. As for Young diagrams, a (semistandard)
skew Young tableau of shape $\la/\mu$ is any filling of the skew Young diagram $\la/\mu$ with
natural numbers which is weakly increasing along rows and strictly increasing down columns. The
corresponding (skew) Schur polynomial $S_{\la/\mu}(x_1,\dots,x_n)$ is defined again by the
right-hand side of Eq.~\eqref{skdef}, where the sum is now over skew Young tableaux of shape
$\la/\mu$. The Jacobi--Trudi formulas for skew Schur polynomials akin to~\eqref{skeh} are
\begin{align}\label{Slamu}
  S_{\la/\mu}(x_1,\dots,x_n)&=\det\bigl(h_{\la_i-\mu_j-i+j}(x_1,\dots,x_n)\bigr)_{i,j=1}^r\\
                            &=
                              \det\bigl(e_{\la'_i-\mu'_j-i+j}(x_1,\dots,x_n)\bigr)_{i,j=1}^s.\notag
\end{align}
Clearly, a skew Young diagram $\lambda/\mu$ need not be a Young diagram. A particular type of skew
Young diagram which in general is not a Young diagram is a \emph{border strip}, i.e., a
connected\footnote{A skew Young diagram is connected if it is possible to join any two of its
  boxes by a path. A path is a sequence of squares such that any two consecutive squares in the
  sequence share a common side.} skew Young diagram with no $2\times2$ blocks. The height of a
border strip is defined as the number of its rows minus one, and its length as the total number of
of its boxes. We shall use the notation $\langle k_1,\ldots,k_r\rangle$ to refer to the border
strip with $k_i$ boxes in the $i$-th column, numbered from right to left
(cf.~Fig.~\ref{fig.border}), and shall denote by $S_{\langle k_1,\dots,k_r\rangle}$ the
corresponding Schur polynomial. Border strips are closely related to \emph{motifs} in the
description of the spectrum of spin chains of Haldane--Shastry type, as we shall discuss in
Section \ref{sec.motif}. This is due to the connection of these diagrams with the corresponding
skew Schur polynomials labeling the irreducible representations of certain Yangian
algebras~\cite{KKN97,NT98}.

All of the above definitions can be readily extended to the $(m|n)$ supersymmetric case by
suitably adapting the definition of semistandard Young tableau. More precisely, given a skew Young
diagram~$\la/\mu$, an $(m|n)$ supersymmetric Young tableau of shape $\la/\mu$ is a filling of
$\la/\mu$ with the integers $1,\dots,m+n$ that is:
\begin{enumerate}[(YT1)]
\item Weakly increasing along rows and strictly increasing down columns for integers in $F$.
\item Strictly increasing along rows and weakly increasing down columns for integers in $B$,
\end{enumerate}
where as usual $B=(b_1,\dots,b_m)$, $F=(f_1,\dots,f_n)$ and $B\cup F=\{1,\dots,m+n\}$ (see
Fig.~\ref{fig.SUSYtab} for an example).
The skew super Schur polynomial $S^{(m|n)}_{\la/\mu}(\bx,\by)$, where $\bx:=(x_1,\dots,x_m)$ and
$\by:=(y_1,\dots,y_n)$, associated with a skew Young diagram $\la/\mu$ is defined by
\begin{equation}\label{Smnlamu}
  S^{(m|n)}_{\la/\mu}(\bx,\by)=\sum_Tx_1^{t_{\vphantom{f_1}b_1}}\cdots
  x_m^{t_{\vphantom{f_1}b_m}}y_1^{t_{f_1}}\cdots y_n^{t_{\vphantom{f_1}f_n}},
\end{equation}
where the sum runs over all the tableaux $T$ of shape $\la/\mu$ filled according to the rules
spelled above. We shall be mainly interested in super Schur polynomials associated with border
strips $\langle k_1,\dots,k_r\rangle$, which we shall denote by
$S_{\langle k_1,\dots,k_r\rangle}^{(m|n)}$. The function
$S^{(m|n)}_{\langle k_1,\dots,k_r\rangle}(\bx,\by)$ is a homogeneous polynomial in the variables
$(\bx,\by)$ of degree equal to the length of the associated border strip. It can be conveniently
expressed in terms of the supersymmetric elementary functions $E^{(m|n)}_{k}(\bx,\by)$ introduced
above by the determinantal formula~\cite{HB00,KKN97,BBHS07}
\begin{equation}\label{Skmn}
  S^{(m|n)}_{\langle k_1,\ldots,k_r\rangle}=\begin{vmatrix}
    E^{(m|n)}_{k_r} & E^{(m|n)}_{ k_{r-1}+k_r} & \cdots & E^{(m|n)}_{k_2+\cdots+k_r} & E^{(m|n)}_{k_1+\cdots+k_r} \\
    1 & E^{(m|n)}_{k_{r-1}} & \cdots & E^{(m|n)}_{k_2+\cdots+k_{r-1}} & E^{(m|n)}_{k_1+\cdots+k_{r-1}} \\
    0 & 1 & \cdots & E^{(m|n)}_{k_2+\cdots +k_{r-2}} & E^{(m|n)}_{k_1+\cdots+k_{r-2}} \\
    \vdots & \vdots && \vdots & \vdots\\
    0 & 0 & \cdots & E^{(m|n)}_{k_2} & E^{(m|n)}_{k_1+k_2} \\
    0 & 0 &\cdots & 1 & E^{(m|n)}_{k_1}
  \end{vmatrix}.
\end{equation}
\begin{figure}[t]
  \begin{ytableau}
    1 & 2 & 3& 4& 4 \\
    2 & 3 & 4& 5 \\
    2
  \end{ytableau}
  \caption{$(3|2)$ supersymmetric Young tableau of shape $(5,4,1)$ for the choice $B=\{1,2,3\}$,
    $F=\{4,5\}$.}
  \label{fig.SUSYtab}
\end{figure}%

\section{Generalized partition function}\label{sec.genPF}
Let us now turn back to the study of the partition function of the supersymmetric HS chain of
$BC_N$ type~\eqref{Hchain}. The method we have applied in Section~\ref{sec.PF} for its computation
is a generalization of that used for the $A_{N-1}$ HS chain in Refs.~\cite{FG05,BB06}. In order to
construct a representation of the partition function of the $BC_N$-type chain in terms of (a
variant of) super Schur polynomials, we shall therefore briefly review how this is done in the
$A_{N-1}$ case~\cite{BBHS07,BBH10}.

\subsection{Review of the $A_{N-1}$ case}

The Hamiltonian of the $\su(m|n)$-supersymmetric HS chain of $A_{N-1}$ type can be taken as
\begin{equation}\label{HA}
  H_A=\frac12 \sum_{i<j}\frac{1- S_{ij}}{\sin^2  \xi^{-}_{ij}},\quad
  \xi_i:=\frac{i \pi}{N},\quad i=1,\dots, N,
\end{equation}
where as usual $S_{ij}$ is the $(m|n)$-supersymmetric spin permutation operator~\eqref{spin-perm}.
Its partition function \cite{FG05,BBHS07} is given by
\begin{equation}\label{ZA}
  Z_A(T)=\sum_{\bk\in{\mathcal P}_N}
  \prod_{i=1}^{r}d^{(m|n)}_{k_i}\cdot F_A(q,\bk),
\end{equation}
where
\begin{equation}\label{dispA}
  \cE_A(i):=i(N-i)
\end{equation}
is the $A_{N-1}$-type dispersion relation and $F_A(q,\bk)$ is defined by the right-hand side
of~\eqref{Fq} with $\cE$ replaced by $\cE_A$. Using the properties of the skew super Schur
polynomials introduced above we can define a generalized partition function of the variables $q$,
$\bx$ and $\by$ by
\begin{equation}
  \cZ_A(q;\bx,\by)=\sum_{\bk\in{\mathcal P}_N}\prod_{i=1}^{r}E^{(m|n)}_{k_i}(\bx,\by)\cdot F_A(q,\bk).
\end{equation}
It follows from Eqs.~\eqref{Ed} and~\eqref{ZA} that the partition function of the $A_{N-1}$-type
HS chain can be expressed in terms of the generalized partition function $\cZ_A$ as
\begin{equation}\label{ZZxy}
  Z_A(q) =\cZ_A(q;1^m,1^n). 
\end{equation}
Since the dispersion relation $\cE_A$ is integer valued, the function $F_A(q,\bk)$, and hence the
generalized partition function~$\cZ_A$, is obviously a polynomial in $q$. It can be shown
\cite{BBH10} that the coefficients of the expansion of $\cZ_A$ in powers of $q$ can be expressed
in terms of the skew super Schur polynomials through the remarkable formula
\begin{equation}\label{ZAqxy}
  \cZ_A(q;\bx,\by)
  =\sum_{\bk\in{\mathcal P}_N}
  S^{(m|n)}_{\langle k_1,\ldots, k_r\rangle}(\bx,\by)\,q^{\sum_{i=1}^{r-1}\cE_A(K_i)}.
\end{equation}
One of the aims of this article is to construct the corresponding expression for the $BC_N$ model.

\subsection{The $BC_N$ case}

Let us now turn to the partition function~\eqref{Zk} of the open supersymmetric Haldane-Shastry
spin chain derived in Section~\ref{sec.PF}. As in the $A_{N-1}$ case, we extend this function to a
generalized partition function depending on the variables ($\bx$, $\by$) replacing the
degeneracies $d^{(m|n)}_{k_i}$, $d^{(m_\vep|n_{\vep'})}_{k_i}$ by the corresponding supersymmetric
elementary functions $E^{(m|n)}_{k_i}(\bx,\by)$, $E_{k_i}^{(m_\vep|n_{\vep'})}(\bx,\by)$. We thus
arrive at the following definition of the generalized partition function in the $BC_N$ case:
\begin{multline}\label{gen}
  \cZ(q;\bx,\by):= \sum_{\bk\in {\mathcal P}_N} \bigg[E^{(m_{\vep}|n_{\vep'})}_{k_r}(\bx,\by)
  \\
  +\bigg(E^{(m|n)} _{k_r}(\bx,\by)-E^{(m_{\vep}|n_{\vep'})}_{k_r}(\bx,\by)\bigg)
  q^{\mathcal{E}(N)}\bigg]\prod_{i=1}^{r-1} E^{(m|n)}_{k_i}(\bx,\by)\cdot F(q,\bk).
\end{multline}
From Eq.~\eqref{Ed} it again follows that the partition function of the $\su(m|n)$ supersymmetric
open HS chain~\eqref{Hchain} can be obtained from the generalized partition function~$\cZ$ by
setting $\bx=(1^m)$, $\by=(1^n)$, i.e.,
\[
  Z(q)=\cZ(q;1^m,1^n).
\]

Our aim is to show that this function can be expressed in terms of suitably modified ($BC_N$-type)
skew super Schur polynomials as
\begin{equation}\label{partition}
  \cZ(q;\bx,\by)=\sum_{\bk\in{\mathcal P}_N}\bigg( S^{(m|n)}_{\langle k_1,\ldots,k_r\rangle,0}(\bx,\by) + S^{(m|n)}_{\langle k_1,\ldots,k_r\rangle,1}(\bx,\by)q^{{\mathcal E}(N)} \bigg)q^{\sum_{i=1}^{r-1}{\mathcal E}(K_i)},
\end{equation}
where
\begin{subequations}
  \label{S01}
  \begin{equation}\label{S0}
    S^{(m|n)}_{\langle k_1,\ldots,k_r\rangle,0}=\begin{vmatrix}
      E^{(m_{\vep}|n_{\vep'})}_{k_r} & E^{(m_{\vep}|n_{\vep'})}_{ k_{r-1}+k_r} & \cdots
      & E^{(m_{\vep}|n_{\vep'})}_{k_1+\cdots+k_r} \\
      1 & E^{(m|n)}_{k_{r-1}} & \cdots &   E^{(m|n)}_{k_1+\cdots+k_{r-1}} \\
      \vdots & \vdots && \vdots \\
      0 & 0 &\cdots &   E^{(m|n)}_{k_1} 
    \end{vmatrix}
  \end{equation}
  \begin{multline}\label{S1}
    S^{(m|n)}_{\langle k_1,\ldots,k_r\rangle,1}\\
    =\begin{vmatrix} \scriptstyle E^{(m|n)}_{k_r}-E^{(m_{\vep}|n_{\vep'})}_{k_r} &\scriptstyle
      E^{(m|n)}_{ k_{r-1}+k_r}- E^{(m_{\vep}|n_{\vep'})}_{ k_{r-1}+k_r} & \cdots & \scriptstyle
      E^{(m|n)}_{k_1+\cdots+k_r}
      -E^{(m_{\vep}|n_{\vep'})}_{k_1+\cdots+k_r} \\
      \scriptstyle 1 & \scriptstyle E^{(m|n)}_{k_{r-1}} & \cdots
      &\scriptstyle E^{(m|n)}_{k_1+\cdots+k_{r-1}} \\
      \vdots & \vdots && \vdots  \\
      \scriptstyle 0 &\scriptstyle 0 &\cdots &\scriptstyle E^{(m|n)}_{k_1}
    \end{vmatrix}.
  \end{multline}
\end{subequations}
Expanding the determinants in Eqs.~\eqref{S01} along the first row and taking into account
Eq.~\eqref{Skmn} we obtain the following equivalent expressions for the polynomials
$S^{(m|n)}_{\langle k_1,\dots,k_r\rangle,\al}$ (with $\al=0,1$):
\begin{equation}
  \label{S0alt}
  S^{(m|n)}_{\langle k_1,\dots,k_r\rangle,\al}
  =(-1)^{r-1}f_{N,\al}+\sum_{s=1}^{r-1}
  (-1)^{s-1}f_{N-K_{r-s},\al}S^{(m|n)}_{\langle k_1,\dots,k_{r-s}\rangle},
\end{equation}
with $f_{k,0}:=E^{(m_\vep|n_{\vep'})}_{k}$ and
$f_{k,1}:=E^{(m|n)}_{k}-E^{(m_\vep|n_{\vep'})}_{k}$. Note also that we obviously have
\begin{equation}\label{SS01}
  S^{(m|n)}_{\langle k_1,\ldots,k_r\rangle}=S^{(m|n)}_{\langle k_1,\ldots,k_r\rangle,0}+S^{(m|n)}_{\langle k_1,\ldots,k_r\rangle,1}.
\end{equation}

The proof of Eq.~\eqref{partition} closely follows the argument in \cite{BB06} for the $A_{N-1}$
case. To begin with, we can rewrite Eq.~\eqref{gen} for the generalized partition function as
\begin{equation}\label{Z0Z1}
  \cZ(q;\bx,\by)=\cZ_0(q;\bx,\by)+q^{\cE(N)}\cZ_1(q;\bx,\by),
\end{equation}
where
\begin{equation}\label{Zal0}
  \cZ_\al(q;\bx,\by):=\sum_{\bk\in\cP_N}F(q,\bk)f_{k_r,\al}(\bx,\by)\prod_{i=1}^{r-1}E^{(m|n)}_{k_i}(\bx,\by)
\end{equation}
and $F(q;\bk)$ is given by Eq.~\eqref{Fq}. Expanding the second product in the definition of $F$
we obtain
\[
  F(q,\bk)=\sum_{\mathclap{\al_1,\dots,\al_{N-r}=0}}^1\,(-1)^{\al_1+\cdots+\al_{N-r}}\,
  q^{\sum_{i=1}^{r-1}\cE(K_i)+\sum_{i=1}^{N-r}\al_i\cE(K'_i)}.
\]
For a given partition $\bk=(k_1,\dots,k_r)\in\cP_N$ of length $r$, the numbers
\[
  \{K_1,\dots,K_{r-1},K'_{i_1},\dots,K'_{i_l}\}
\]
(with $1\le i_1<\cdots<i_l\le N-r$ and $l\le N-r$) are clearly the partial sums
\[
  \{\hK_1,\dots,\hK_{s-1}\}
\]
(excluding $\hK_s=N$) of another such partition $\widehat\bk\in\cP_N$ of length $s=r+l$ finer
than $\bk$. We can thus rewrite $\cZ_\al$ as
\begin{equation}\label{Zal}
  \cZ_\al(q;\bx,\by)=\sum_{\bk
    \in\mathcal{P}_N}\fS_{\bk,\al}(\bx,\by)q^{\sum_{i=1}^{r-1}\mathcal{E}(K_i)},
\end{equation}
where the coefficients $\fS_{\bk}(\bx,\by)$ are to be determined. From the previous discussion it
is clear that the only partitions $\tbk=(\tk_1,\dots,\tk_s)\in\cP_N$ in Eq.~\eqref{Zal0} which
contribute to the term $q^{\sum_{i=1}^{r-1}\cE(K_i)}$ in the previous sum are those coarser than
$\bk$, i.e., such that
\[
  \{\tK_1,\dots,\tK_{s-1}\}\subset\{K_1,\dots,K_{r-1}\}\,.
\]
Defining the integers $L_1<\dots<L_{s-1}$ by $\tK_i=K_{L_i}$, and noting that
$\tk_i=\tK_i-\tK_{i-1}$ (with $\tK_0:=0$) for $i=1,\dots,s-1$ and
$\tk_s=N-\tK_{s-1}=N-K_{L_{s-1}}$, we thus obtain
\begin{multline}\label{fsbk0}
  \fS_{\bk,\al}
  =(-1)^{r-1}f_{N,\al}\\
  +\sum_{s=2}^{r}\,\sum_{1\le L_1<\cdots<L_{s-1}\le
    r-1}(-1)^{r-s}f_{N-K_{L_{s-1}},\al}\prod_{i=1}^{s-1}E^{(m|n)}_{K_{L_i}-K_{L_{i-1}}},
\end{multline}
where the first term corresponds to the partition $\tbk=(N)$ of length $s=1$. We next define the
integers $\ell_i:=L_i-L_{i-1}\ge1$ (with $i=1,\dots,s-1$) and $L_0:=0$, in terms of which
$L_i=\sum_{j=1}^i\ell_j$. Calling $p=r-L_{s-1}\ge1$, it follows that
$\bell:=(\ell_1,\dots,\ell_{s-1})$ belongs to the set $\cP_{r-p}(s-1)$ of partitions (taking order
into account) of the integer $r-p$. Since $p=r-\sum_{i=1}^{s-1}\ell_i\le r-s+1$, we can rewrite
Eq.~\eqref{fsbk0} as
\[
  \fS_{\bk,\al}=(-1)^{r-1}f_{N,\al}+\sum_{s=2}^{r}\sum_{p=1}^{r-s+1}\sum_{\bell\in\cP_{r-p}(s-1)}
  (-1)^{r-s}f_{N-K_{r-p},\al}\prod_{i=1}^{s-1}E^{(m|n)}_{K_{L_i}-K_{L_{i-1}}}.
\]
Exchanging the sums over $s$ and $p$ and setting $s=j+1$ we obtain
\begin{align*}
  \fS_{\bk,\al}&=(-1)^{r-1}f_{N,\al}
  \\
               &\hphantom{(-1)^{r-1}}+\sum_{p=1}^{r-1}(-1)^{p-1}f_{N-K_{r-p},\al}
                 \sum_{j=1}^{r-p}(-1)^{r-p-j}\sum_{\bell\in\cP_{r-p}(j)}
                 \prod_{i=1}^{j}E^{(m|n)}_{K_{L_i}-K_{L_{i-1}}}.
\end{align*}
Recalling the identity
\[
  \sum_{j=1}^{r-p}(-1)^{r-p-j}\sum_{\bell\in\cP_{r-p}(j)}
  \prod_{i=1}^{j}E^{(m|n)}_{K_{L_i}-K_{L_{i-1}}}=S^{(m|n)}_{\langle k_1,\dots,k_{r-p}\rangle}
\]
(cf.~Eq.~(3.25) in Ref.~\cite{BBHS07}) and comparing with Eq.~\eqref{S0alt} we conclude that
\[
  \fS_{\bk,\al}=S^{(m|n)}_{\langle k_1,\dots,k_r\rangle,\al}.
\]
Equation~\eqref{partition} then follows immediately from the latter relation
and~\eqref{Z0Z1}-\eqref{Zal}.

\begin{rem}
  Following Ref.~\cite{BS20}, equation~\eqref{partition} can be written as
  \begin{equation}
    \label{ZSchurq}
    \cZ(q;\bx,\by)=\sum_{\bk\in{\mathcal P}_N}S^{(m|n)}_{\langle k_1,\ldots,k_r\rangle}(q;\bx,\by)q^{\sum_{i=1}^{r-1}{\mathcal E}(K_i)},
  \end{equation}
  where
  \begin{subequations}\label{Schrqeqs}
    \begin{equation}
      \label{Schurq}
      S^{(m|n)}_{\langle k_1,\ldots,k_r\rangle}(q;\bx,\by)=\sum_{l=0}^NS^{(m|n)}_{\langle  k_1,\ldots,k_r|l\rangle}
      (\bx,\by)q^{\cE(l)}
    \end{equation}
    with
    \begin{equation}
      \label{Schurql}
      S^{(m|n)}_{\langle  k_1,\ldots,k_r|l\rangle}(\bx,\by)=\de_{l0}S^{(m|n)}_{\langle
        k_1,\ldots,k_r\rangle,0}(\bx,\by)+\de_{lN}S^{(m|n)}_{\langle
        k_1,\ldots,k_r\rangle,1}(\bx,\by)\,.
    \end{equation}
  \end{subequations}
  We see that Eq.~\eqref{ZSchurq} is formally the analogue of Eq.~\eqref{ZAqxy}, with the skew
  super Schur polynomials $S^{(m|n)}_{\langle k_1,\ldots,k_r\rangle}(\bx,\by)$ replaced by their
  $q$-deformed versions~\eqref{Schurq}. (Note, in this respect, that the dispersion relation of
  the Polychronakos (rational) chain of $BC_N$ type discussed in Ref.~\cite{BS20} is simply
  $\cE(x)=x$.) A major difference between Eq.~\eqref{Schurql} and its counterpart for the rational
  chain studied in Ref.~\cite{BS20} is the fact that in our case the only nonvanishing polynomials
  $S^{(m|n)}_{\langle k_1,\ldots,k_r|l\rangle}(\bx,\by)$ are the first ($l=0$) and the last one
  ($l=N$), whereas for the rational chain all of the corresponding polynomials are in general
  nonzero. An important consequence of this fact is that the ``branching'' of the spectrum of the
  Polychronakos chain of $BC_N$ type is far greater than is the case for the present model, as we
  shall explain in Remark~\ref{rem.branching} below.
\end{rem}
\section{$BC_N$-type motifs and border strips}\label{sec.motif}

In this section we shall take advantage of the explicit formula~\eqref{partition} for the
$BC_N$-type generalized partition function~$\cZ(q;\bx,\by)$ to show that the spectrum of the HS
chain~\eqref{Hchain} coincides with that of a vertex model with appropriate interactions between
consecutive vertices plus an additional boundary term. This will lead to a description of the
chain's spectrum in terms of a novel $BC_N$-type version of Haldane's
motifs~\cite{HHTBP92,Ha93,HB00,BBHS07}. As before, it will prove convenient to start by reviewing
the motif-based description of the spectrum of the $A_{N-1}$-type HS chain.

\subsection{Review of the $A_{N-1}$ case}

From Eq.~\eqref{ZAqxy} it follows that the partition function of the $\su(m|n)$ supersymmetric HS
chain of type $A_{N-1}$ can be expressed as
\begin{equation}
  \label{Zamotifs}
  Z_A=\sum_{\bk\in\cP_N}d_A(\bk)\,q^{\sum_{i=1}^{r-1}\cE_A(K_i)},
\end{equation}
where by Eq.~\eqref{Smnlamu}
\begin{equation}
  d_A(\bk)=S^{(m|n)}_{\langle k_1,\ldots, k_r\rangle}(1^m,1^n).
  \label{dak}
\end{equation}
This shows that the spectrum of $H_A$ consists of the numbers (nonnegative integers)
\begin{equation}\label{specHA}
  E_A(\bk)=\sum_{i=1}^{r-1}\cE_A(K_i),\qquad\bk=(k_1,\dots,k_r)\in\cP_N,
\end{equation}
each of which possesses an intrinsic degeneracy $d_A(\bk)$. Moreover, by Eq.~\eqref{Smnlamu}
$d_A(\bk)$ is the number of $(m|n)$ supersymmetric skew Young tableaux corresponding to the border
strip~$\langle k_1,\dots,k_r\rangle$, i.e., the number of fillings of the latter border strip with
the integers $\{1,\dots,m+n\}=B\cup F$ consistent with rules (YT1)-(YT2) in Section~\ref{sec.Schur}.
These facts make it possible to find a motif-based description of the spectrum of the
supersymmetric HS chain of type $A_{N-1}$ as follows.

To each tableau corresponding to the border strip $\langle k_1,\dots,k_r\rangle$ we associate a
\emph{bond vector} $\bsv=(s_1,\dots,s_N)\in(B\cup F)^N$ whose components are the numbers filling
the tableau read from right to left and top to bottom. This obviously establishes a one-to-one
correspondence between tableaux associated with a border strip and allowed bond vectors, where a
bond vector is said to be allowed if its corresponding tableau satisfies rules (YT1)-(YT2). It is
apparent that the energy~$E_{\bk,A}$ associated to a given border strip
$\langle k_1,\dots,k_r\rangle$ is given by
\begin{equation}\label{EAbk}
  E_{\bk,A}=\sum_{j=1}^{N-1}\de_j\cE_A(j)\,,
\end{equation}
where $\de_j=1$ if $j\in\{K_1,\dots,K_{r-1}\}$ and $\de_j=0$ otherwise. The vector
$\bde:=(\de_1,\dots,\de_{N-1})\in\{0,1\}^{N-1}$ is the \emph{motif} corresponding to the border
strip $\langle k_1,\dots,k_r\rangle$. Note that there is also a one-to-one correspondence between
border strips and motifs, since given a motif $\bde$ the associated border strip can be
constructed by starting with one empty box and successively adding a box to the left of the $i$-th
box (respectively below the $i$-th box) provided that $\de_i=1$ (resp.~$\de_i=0$). Again, we shall
say that the motif $\bde$ is allowed if its corresponding border strip admits at least one tableau
consistent with rules (YT1)-(YT2) above. It is easy to see that in the truly supersymmetric case
$mn\ne0$ all motifs are allowed, whereas in the purely bosonic case $n=0$ (resp.~purely fermionic
case $m=0$) the only allowed motifs are those containing no sequence with $m$ or more $1$'s
(resp.~$n$ or more $0$'s).

Given an allowed motif $\bde$, the intrinsic degeneracy of its energy~\eqref{EAbk} is given by the
number of tableaux corresponding to the border strip $\langle k_1,\dots,k_r\rangle$ associated
with $\bde$, or equivalently of bond vectors allowed for this border strip. Since the $1$'s in the
motif $\bde$ occupy by construction the positions labeled by the partial sums
$\{K_1,\dots,K_{r-1}\}$ (called \emph{rapidities} in the literature) of the partition
$\bk\in\cP_N$ corresponding to the border strip~$\langle k_1,\dots,k_r\rangle$, a moment's
reflection shows that a bond vector $\bsv=(s_1,\dots,s_N)$ is allowed for the border strip
$\langle k_1,\dots,k_r\rangle$ constructed from the motif $\bde$ if and only if
$\de_i=\de(s_i,s_{i+1})$ for $i=1,\dots,N-1$, where the function $\de:(B\cup F)^2\to\{0,1\}$ is
defined by
\begin{equation}
  \label{dedefA}
  \de(s,t)=
  \begin{cases}
    0,\quad &s<t \text{ or }s=t\in B,\\
    1, &s>t \text{ or }s=t\in F.
  \end{cases}
\end{equation}
We conclude that the spectrum of the $\su(m|n)$ supersymmetric HS chain~\eqref{HA} (with the
correct degeneracy for each level) can be generated through the formula
\begin{equation}
  \label{specmotif}
  E_A(\bsv)=\sum_{i=1}^{N-1}\de(s_i,s_{i+1})\cE_A(i)\,,
\end{equation}
where the bond vector $\bsv$ runs over the set $(B\cup F)^N$.

Equation~\eqref{specmotif} admits an obvious interpretation as the energy function of a classical
vertex model with $N+1$ vertices and $N$ bonds, where each bond can be in one of $m+n$ possible
states, $m$ of which are of ``bosonic'' and the remaining $n$ of ``fermionic'' type. Indeed, it
suffices to assign the energy $\de(s_i,s_{i+1})\cE_A(i)$ to the $i$-th bond if $i=1,\dots,N-1$,
and zero energy to the last ($N$-th) bond. The vertex model's partition function can thus be
written as
\[
  Z_A^V(q)=\sum_{\bsv\in(B\cup F)^N}q^{E_A(\bsv)}.
\]
Note that, by construction, we have
\[
  Z_A^V(q)=Z_A(q).
\]

\subsection{The $BC_N$ case}

Let us now turn back to the $BC_N$ case. To begin with, setting
\begin{equation}\label{Stil}
  \tS^{(m|n)}_{\langle k_1,\ldots,k_r\rangle}(\bx,\by)=\begin{cases}
    S^{(m|n)}_{\langle k_1,\ldots,k_{r-1},k_r-1\rangle,0}(\bx,\by),&k_r>1\\[10pt]
    S^{(m|n)}_{\langle k_1,\ldots,k_{r-2},k_{r-1}\rangle,1}(\bx,\by),&k_r=1
  \end{cases}
\end{equation}
we can more conveniently rewrite Eq.~\eqref{partition} as
\begin{equation}\label{partition2}
  \cZ(q;\bx,\by)=\sum_{\bk\in{\mathcal P}_{N+1}}
  \tS^{(m|n)}_{\langle k_1,\ldots,k_r\rangle}(\bx,\by)\,q^{\sum_{i=1}^{r-1}{\mathcal E}(K_i)}.
\end{equation}
It should be stressed that the border strip $\langle k_1,\ldots,k_r\rangle$ in the LHS of
Eq.~\eqref{Stil} corresponds to a partition $\bk=(k_1,\dots,k_r)$ of $N+1$ of length $r$, whereas
the border strips ${\langle k_1,\ldots,k_{r-1},k_r-1\rangle}$ and
$\langle k_1,\ldots,k_{r-1}\rangle$ in the RHS correspond to partitions of $N$ with respective
lengths $r$ and $r-1$.

Equation~\eqref{partition2} again entails that the partition function of open HS
chain~\eqref{Hchain} can be expressed in a more compact way as
\begin{equation}
  \label{ZBS}
  Z(q)=\sum_{\bk\in{\mathcal P}_{N+1}} d(\bk)\,q^{\sum_{i=1}^{r-1}{\mathcal E}(K_i)},
\end{equation}
with
\[
  d(\bk)=\tS^{(m|n)}_{\langle k_1,\ldots,k_r\rangle}(1^m,1^n)=\begin{cases}
    S^{(m|n)}_{\langle k_1,\ldots,k_{r-1},k_r-1\rangle,0}(1^m,1^n)=:d_0(\bk),&k_r>1\\[10pt]
    S^{(m|n)}_{\langle k_1,\ldots,k_{r-2},k_{r-1}\rangle,1}(1^m,1^n)=:d_1(\bk)&k_r=1.
  \end{cases}
\]
By Eq.~\eqref{Ed}, $d_\al(\bk)$ can be obtained replacing $E^{(m|n)}_k$ and
$E_k^{(m_\vep|n_{\vep'})}$ in Eqs.~\eqref{S01} by $d^{(m|n)}_k$ and $d_k^{(m_\vep|n_{\vep'})}$,
respectively.

To get a better understanding of the latter formulas, it is useful to consider them in the two
simple cases $\bk=(N+1)$ and $\bk=(N,1)$. In the first case, from Eq.~\eqref{S0} with
$(\bx,\by)=(1^m,1^n)$ it follows that
\[
  d(\bk)=d_0(\bk)=d_N^{(m_\vep|n_{\vep'})}.
\]
Hence the degeneracy corresponding to the single column border strip $\langle N+1\rangle$ is equal
to the number of $(m_\vep|n_{\vep'})$ skew Young tableaux of shape $\langle N\rangle$. If we order
the spin variables so that
\begin{equation}
  \label{order}
  \{b_1,\dots,b_{m_\vep}\}\cup\{f_1,\dots,f_{n_{\vep'}}\}=\{1,\dots,m_\vep+n_{\vep'}\},\quad
  b_{m_\vep}=m_\vep+n_{\vep'},
\end{equation}
it is easy to convince oneself that one can express $d(\bk)$ as the number of $(m|n)$
supersymmetric skew Young tableaux of shape $\langle N+1\rangle$ with the last box filled with the
integer $*:=m_\vep+n_{\vep'}$, which is by construction of \emph{bosonic} type. We shall
symbolically denote the shape of these tableaux by
\[
 \ytableaushort{{},{\raise-2pt\hbox{$\vdots$}},{},{\raise1pt\hbox{$*$}}}
\]
Equivalently, $d(\bk)$ is given by the number of bond vectors $\bsv=(s_1,\dots,s_{N+1})$
corresponding to the partition $\bk=(N+1)$ according to the rules (YT1)-(YT2) in
Section~\ref{sec.Schur}, with $s_{N+1}=*$ fixed.
\begin{rem}\label{rem.mepone}
  When $m_\vep=0$ and $n_{\vep'}>0$ (i.e., for $m=0$ or $m=n=\vep'=-\vep=1$), we have
  $*=n_{\vep'}$, $B_\vep=\emptyset$, and $B_\vep\cup F_{\vep'}=F_{\vep'}$. It should be clear from
  the above discussion that in this case we should still regard the $*$ symbol in the last box as
  \emph{bosonic}, even if $n_{\vep'}=f_{n_{\vep'}}$ is of fermionic type otherwise. Indeed, in
  this case $n_{\vep'}$ is \emph{allowed} in the box above the last (starred) one.
\end{rem}

Likewise, for the partition $\bk=(N,1)\in\cP_{N+1}$ by Eq.~\eqref{S1} with
$(\bx,\by)=(1^m,1^n)$ we have
\[
  d(\bk)=d_1(\bk)=
    d_N^{(m|n)}-d_N^{(m_\vep|n_{\vep'})}.
\]
Thus $d(\bk)$ equals the difference between the number of $(m|n)$ and $(m_\vep|n_{\vep'})$
supersymmetric tableaux corresponding to the single column border strip $\langle N\rangle$. This
is evidently equal to the number of tableaux of shape $\langle N\rangle$ containing at least one
entry greater than $m_\vep+n_{\vep'}$, which in turn (since skew Young tableaux are non-decreasing
down columns) coincides with the number of such tableaux whose last entry is greater than
$m_\vep+n_{\vep'}$. Since $*=m_\vep+n_{\vep'}=b_{m_\vep}$ is of bosonic type, this is the same as
the number of tableau of shape $\langle N,1\rangle$ whose last box is filled by $*$. We shall
again indicate the shape of these tableaux by the diagram
\[
 \ytableaushort{\none{},\none{\raise-2pt\hbox{$\vdots$}},{\raise1pt\hbox{$*$}}{}}
\]
Equivalently, $d(\bk)$ is given by the number of bond vectors $\bsv=(s_1,\dots,s_{N+1})$
corresponding to the partition $(N,1)$ according to the rules (YT1)-(YT2), again with $s_{N+1}=*$
fixed.
\begin{rem}\label{rem.meptwo}
  As in the previous example, when $m_\vep=0$ and $n_{\vep'}>0$ the symbol $*=n_{\vep'}$ should be
  regarded as \emph{bosonic} even if $n_{\vep'}=f_{n_{\vep'}}$ is fermionic in this case, since
  from the preceding discussion it follows that $n_{\vep'}$ is \emph{not} allowed in the box to
  the right of the last (starred).
\end{rem}
The above considerations suggest that in \emph{all} cases the degeneracy associated with a
partition $\bk\in\cP_{N+1}$ is the number of $(m|n)$ supersymmetric skew Young tableau of shape
$\langle k_1,\dots,k_r\rangle$ with the last (leftmost and lowermost) box filled with
$*=m_\vep+n_{\vep'}$, regarded always as bosonic. We shall prove below that this is indeed the
case. More precisely:

\begin{enumerate}[(R1)]
\item The eigenvalues of the open $\su(m|n)$ HS chain~\eqref{Hchain} are labeled by the partitions
  (with order taken into account) $\bk=(k_1,\dots,k_r)$ of the integer $N+1$ according to the
  formula.
  \[
    E(\bk)=\sum_{i=1}^{r-1}\cE(K_i)\,,
  \]
  where the dispersion relation $\cE$ is defined by Eq.~\eqref{disprel}.
\item The intrinsic degeneracy $d(\bk)$ of the eigenvalue $E(\bk)$ (which could possibly be equal
  to zero) coincides with the number of $(m|n)$ supersymmetric skew Young tableaux of shape
  $\langle k_1,\dots, k_r\rangle$ of length $N+1$ whose last (i.e, lowermost and leftmost) box is
  filled with $*=m_\vep+n_{\vep'}$, regarded always as bosonic.
\end{enumerate}
\begin{rem}
  It is of course understood that the spin variables must be chosen according to the
  convention~\eqref{order} (with the proviso mentioned in Remark~\ref{rem.mepone} when $m_\vep=0$
  and $n_{\vep'}>0$), which we shall tacitly follow in the sequel. There are obviously other
  conventions yielding the same rule for the degeneracy $d(\bk)$. For instance, we could have
  equivalently set $*=m_\vep+n_{\vep'}+1=f_{n_{\vep'}+1}$, regarded as fermionic even when
  $m_\vep>0$ and $n_{\vep'}=0$. Alternatively, we could have defined $*=m_\vep+n_{\vep'}+1/2$,
  which has the advantage of not requiring a special proviso when $m_\vep$ or $n_{\vep'}$ vanish.
  It should also be noted that $d(\bk)$ could be zero for some partitions $\bk\in\cP_{N+1}$ (even
  in the truly supersymmetric case $mn\ne0$), in which case~$\sum_{i=1}^{r-1}\cE(K_i)$ is
  \emph{not} an eigenvalue\footnote{Unless, of course,
    $\sum_{i=1}^{r-1}\cE(K_i)=\sum_{i=1}^{\tilde r-1}\cE(\tK_i)$ for some other
    partition~$\tbk:=(\tk_1,\dots,\tk_{\tilde r})\in\cP_{N+1}$ with
    $d(\tbk)>0$.\vphantom{$T^{T^{T^T}}$}} of the chain~\eqref{Hchain}.
\end{rem}

Before proving the above two rules, we shall briefly outline some of its main consequences. First
of all, as in the $A_{N-1}$ case, from (R1)-(R2) above it follows that the spectrum of the
supersymmetric HS chain of $BC_N$ type can be equivalently described in terms of ``starred''
border strips (i.e, with the last boxed filled by $*$) and motifs, where now the motifs have
length $N$ instead of $N-1$. In other words, the eigenvalues of the Hamiltonian~\eqref{Hchain} can
be generated by the formula ---akin to its~$A_{N-1}$ counterpart~\eqref{EAbk}---
\begin{equation}\label{EdB}
  E_{\bde}=\sum_{i=1}^{N}\de_i\cE(i),\qquad \bde:=(\de_1,\dots,\de_N)\in\{0,1\}^N.
\end{equation}
The degeneracy of the eigenvalue~$E_{\bde}$ (which can possibly be zero) is given by the number of
$(m|n)$ supersymmetric \emph{starred} tableaux having as shape the border strip corresponding to
the motif $\bde$. We stress that the rule for filling the tableaux is exactly the \emph{same} as
in the $A_{N-1}$ case, i.e., is given by conditions (YT1)-(YT2) in Section~\ref{sec.Schur}. The
only differences with the latter case are that i)~the tableaux now have one extra box (i.e., they
are of length $N+1$), and ii)~the last box must be filled by $*=m_\vep+n_{\vep'}$, regarded always
as bosonic.

Just as in the $A_{N-1}$ case, the previous description of the spectrum of the $BC_N$
chain~\eqref{Hchain} can be reformulated in the framework of classical vertex models. Indeed, the
spectrum of the chain~\eqref{Hchain} can be equivalently generated using bond vectors
$(s_1,\dots,s_{N+1})\in(B\cup F)^{N+1}$ with $s_{N+1}=*$ by setting
\begin{equation}\label{EsB}
  E_{\bsv}=\sum_{i=1}^{N}\de(s_i,s_{i+1})\cE(i),
\end{equation}
where $\de:(B\cup F)^2\to\{0,1\}$ is defined exactly as in the~$A_{N-1}$ case
(Eq.~\eqref{dedefA}). The latter formula can of course be interpreted as the energy function of a
classical vertex model with $N+2$ vertices and $N+1$ bonds each of which can be in one of $m+n$
possible states, $m$ of which are bosonic and $n$ fermionic, with the following two restrictions:
i)~the last bond has zero energy, and ii)~the bond before the last is always in the state
$*=m_\vep+n_{\vep'}$, regarded as bosonic.

We shall now provide a complete proof of rules (R1)-(R2) above. The proof will be based on an
alternative recursion relation satisfied by the $BC_N$-type super Schur polynomials
$S^{(m|n)}_{\langle k_1,\dots,k_r\rangle,\al}$ obtained expanding the determinants in
Eq.~\eqref{S01} along their first column, namely
\begin{equation}\label{recS}
  S^{(m|n)}_{\langle k_1,\dots,k_r\rangle,\al}=f_{k_r,\al}S^{(m|n)}_{\langle
    k_1,\dots,k_{r-1}\rangle}-S^{(m|n)}_{\langle k_1,\dots,k_{r-2},k_{r-1}+k_r\rangle,\al}.
\end{equation}

First of all, it is clear from Eq.~\eqref{partition2} that the eigenvalues of the
Hamiltonian~\eqref{Hchain} can only be the numbers $\sum_{i=1}^{r-1}\cE(K_i)$, where $\bk$ is a
partition of $N+1$ of length $r$. This establishes the first rule. The second one will be proved
by induction on the number of columns $r'$ of the border strip corresponding to a given partition
of the integer $N+1$ with the last box removed. The two examples presented above then show that
the rules (R1)-(R2) are valid for $r'=1$. Assume, therefore, that they hold for partitions of
$N+1$ with $r'\le \rho$, and consider a partition with $r'=\rho+1$. Suppose, first, that this
partition is of the type $\bk=(k_1,\dots,k_r)$ with $k_r>1$, so that $d(\bk)=d_0(\bk)$ and
$r=r'=\rho+1$. Evaluating the identity~\eqref{recS} with $\al=0$ at the point
$(\bx,\by)=(1^m,1^n)$ we obtain the recursion relation
\begin{align}
  \label{recd0}
  d(\bk)&=d_0(\bk)=S^{(m|n)}_{\langle k_1,\dots,k_r-1\rangle,0}(1^m,1^n)\\
        &=d_A(k_1,\dots,k_{r-1})d^{(m_\vep|n_{\vep'})}_{k_r-1}-d_0(k_1,\dots,k_{r-2},k_{r-1}+k_r-1).
          \notag
\end{align}
The first term in the RHS is the number of fillings of the border strip
$\langle k_1,\dots,k_{r-1}\rangle$ according to the $(m|n)$ supersymmetric rules (YT1)-(YT2) in
Section~\ref{sec.Schur} times \emph{all} possible fillings of a single column of height $k_r-1$
using only the integers $\{1,\dots,m_\vep+n_{\vep'}\}$. By the induction hypothesis, the second
term counts the number of fillings of the border strip
$\langle k_1,\dots,k_{r-2},k_{r-1}+k_r-1\rangle$ according to rule (R2), i.e., such that the last
box is filled only with the integers~{$\{1,\dots,m_\vep+n_{\vep'}\}$}. We now use the following
elementary identity involving border strips: \definecolor{lightgray}{gray}{0.85}
\begin{equation}\label{tabprod}\abovedisplayskip=0pt
  \ytableausetup{centertableaux} \ytableaushort[*(lightgray)]{{},{\raise-2pt\hbox{$\vdots$}},{}}
  \en\times\overset{^{\hphantom{}\ytableaushort{\none}}{}^{\displaystyle\iddots}}%
  {\ytableaushort{{},{\raise-2pt\hbox{$\vdots$}},{}}}
  =\en\overset{^{\hphantom{}\ytableaushort{\none\none}}{}^{\displaystyle\iddots}}%
  {\ytableaushort{\none{},\none{\raise-2pt\hbox{$\vdots$}},{*(lightgray)}{},
      {*(lightgray)\raise-2pt\hbox{$\vdots$}}\none,{*(lightgray)}}}
  \en+\overset{^{\hphantom{}\ytableaushort{\none}}{}^{\displaystyle\iddots}}%
  {\ytableaushort[*(lightgray)]{{},{\raise-2pt\hbox{$\vdots$}},{},{},{\raise-2pt\hbox{$\vdots$}},{}}}
\end{equation}
where each border strip represents the total number of $(m|n)$ supersymmetric Young tableaux
associated with it, and the shaded columns are filled using only the integers
$\{1,\dots,m_\vep+n_{\vep'}\}$. Equation~\eqref{recd0} can thus be symbolically expressed as
\[\abovedisplayskip=0pt
  d(\bk)=\en
  \ytableausetup{centertableaux} \ytableaushort[*(lightgray)]{{},{\raise-2pt\hbox{$\vdots$}},{}}
  \en\times\overset{^{\hphantom{}\ytableaushort{\none}}{}^{\displaystyle\iddots}}%
  {\ytableaushort{{},{\raise-2pt\hbox{$\vdots$}},{}}}\en-
\en\overset{^{\hphantom{}\ytableaushort{\none}}{}^{\displaystyle\iddots}}%
{\ytableaushort[*(lightgray)]{{},{\raise-2pt\hbox{$\vdots$}},{},{},{\raise-2pt\hbox{$\vdots$}},{}}}
=\overset{^{\hphantom{}\ytableaushort{\none\none}}{}^{\displaystyle\iddots}}%
  {\ytableaushort{\none{},\none{\raise-2pt\hbox{$\vdots$}},{*(lightgray)}{},
      {*(lightgray)\raise-2pt\hbox{$\vdots$}}\none,{*(lightgray)}}}
  =\overset{^{\hphantom{}\ytableaushort{\none\none}}{}^{\displaystyle\iddots}}%
  {\ytableaushort{\none{},\none{\raise-2pt\hbox{$\vdots$}},{}{},
      {\raise-2pt\hbox{$\vdots$}}\none,{}\none,{\raise1pt\hbox{$*$}}\none}}
\]
Thus in this case $d(\bk)$ is equal to the number of fillings of the border strip
$\langle k_1,\dots,k_r\rangle$ according to the rule (R2) (i.e., filling the last box with the
integer $m_\vep+n_{\vep'}$), as claimed.

Consider next a partition $\bk=(k_1,\dots,k_r)$ of $N+1$ with $r'=\rho+1$ and $k_r=1$, so that
$d(\bk)=d_1(\bk)$ and $r'=r-1=\rho+1$. Evaluating the identity~\eqref{recS} with $\al=1$ at the
point $(\bx,\by)=(1^m,1^n)$ we obtain the recursion relation
\begin{align}
  \label{recd1}
  d(\bk)&=d_1(\bk)=S^{(m|n)}_{\langle k_1,\dots,k_{r-1}\rangle,1}(1^m,1^n)\\
        &=\Big(d^{(m|n)}_{k_{r-1}}-d^{(m_\vep|n_{\vep'})}_{k_{r-1}}\Big)
          d_A(k_1,\dots,k_{r-2})-d_1(k_1,\dots,k_{r-3},k_{r-2}+k_{r-1}).
          \notag
\end{align}
The term in parentheses in the RHS is equal to the number of fillings of the single column of
length $k_{r-1}$ whose last box contains only integers greater than $m_\vep+n_{\vep'}$. By the
induction hypothesis, the last term represents the number of fillings of the border strip
$\langle k_1,\dots,k_{r-3},k_{r-2}+k_{r-1}\rangle$ whose last box is filled with integers also
greater than $m_\vep+n_{\vep'}$. We can thus symbolically express Eq.~\eqref{recd1} as
 \[\abovedisplayskip=0pt \abovedisplayskip=0pt
   d(\bk)=\en \ytableausetup{centertableaux}
   \ytableaushort{{},{\raise-2pt\hbox{$\vdots$}},{*(lightgray)}}
   \en\times\overset{^{\hphantom{}\ytableaushort{\none}}{}^{\displaystyle\iddots}}%
   {\ytableaushort{{},{\raise-2pt\hbox{$\vdots$}},{}}}\en-
   \en\overset{^{\hphantom{}\ytableaushort{\none}}{}^{\displaystyle\iddots}}%
   {\ytableaushort{{},{\raise-2pt\hbox{$\vdots$}},{},{},{\raise-2pt\hbox{$\vdots$}},{*(lightgray)}}}
\]
where the gray box is filled only with integers greater than $m_\vep+n_{\vep'}$. By the elementary
identity
 \[\abovedisplayskip=0pt
   \ytableausetup{centertableaux} \ytableaushort{{},{\raise-2pt\hbox{$\vdots$}},{*(lightgray)}}
   \en\times\overset{^{\hphantom{}\ytableaushort{\none}}{}^{\displaystyle\iddots}}%
   {\ytableaushort{{},{\raise-2pt\hbox{$\vdots$}},{}}}
   =\en\overset{^{\hphantom{}\ytableaushort{\none\none}}{}^{\displaystyle\iddots}}%
   {\ytableaushort{\none{},\none{\raise-2pt\hbox{$\vdots$}},{}{},
       {\raise-2pt\hbox{$\vdots$}}\none,{*(lightgray)}}}
   \en+\overset{^{\hphantom{}\ytableaushort{\none}}{}^{\displaystyle\iddots}}%
   {\ytableaushort{{},{\raise-2pt\hbox{$\vdots$}},{},{},{\raise-2pt\hbox{$\vdots$}},{*(lightgray)}}}
 \]
 we conclude that in this case
 \[\abovedisplayskip=0pt
   d(\bk)\en=\overset{^{\hphantom{}\ytableaushort{\none\none}}{}^{\displaystyle\iddots}}%
   {\ytableaushort{\none{},\none{\raise-2pt\hbox{$\vdots$}},{}{},
       {\raise-2pt\hbox{$\vdots$}}\none,{*(lightgray)}}}\en=\qquad\mathclap{\overset{^{\hphantom{}\ytableaushort{\none\none\none}}{}^{\displaystyle\iddots}}%
     {\ytableaushort{\none\none{},\none\none{\raise-2pt\hbox{$\vdots$}},\none{}{},
         \none{\raise-2pt\hbox{$\vdots$}}\none,{\raise1pt\hbox{$*$}}{}}}}
   \]
   as claimed. This completes the proof of rules (R1)-(R2) above.

\begin{rem}\label{rem.Scomb}
  A similar argument can be used to find the following combinatorial expression for the
  $BC_N$-type super Schur polynomials $S^{(m|n)}_{\langle k_1,\dots,k_r\rangle,\al}$:
  \begin{equation}
    S^{(m|n)}_{\langle k_1,\dots,k_r\rangle,\al}=\sum_{T\in\cT_\al}x_1^{t_{\vphantom{f_1}b_1}}\cdots
    x_m^{t_{\vphantom{f_1}b_m}}y_1^{t_{f_1}}\cdots y_n^{t_{\vphantom{f_1}f_n}},
  \end{equation}
  where $\cT_0$ (respectively $\cT_1$) denotes the set of all supersymmetric tableaux of
  shape~$\langle k_1,\dots,k_r\rangle$ whose last box is filled by an integer
  $\le m_\vep+n_{\vep'}$ (respectively~$>m_\vep+n_{\vep'}$). Indeed, the formula is clearly true
  for $r=1$, and it can be easily proved by induction on $r$ using the recursion
  relation~\eqref{recS}.
\end{rem}
\begin{rem}\label{rem.branching}
  Setting $\bx=(1^m)$, $\by=(1^n)$ in Eq.~\eqref{partition} or~\eqref{ZSchurq} we obtain the
  following alternative formula for the partition function of the $\su(m|n)$ HS chain of $BC_N$
  type:
  \begin{equation}\label{Zqalt}
    Z(q)=\sum_{\bk\in\cP_N}\big(d_0(\bk)+d_1(\bk)q^{\cE(N)}\big)q^{\sum_{i=1}^{r-1}\cE(K_i)}\,,
  \end{equation}
  where
  \[
    d_0(\bk)=S^{(m|n)}_{\langle k_1,\dots,k_r|0\rangle}(1^m,1^n)\,,\qquad
    d_1(\bk)=S^{(m|n)}_{\langle k_1,\dots,k_r|N\rangle}(1^m,1^n)\,.
  \]
  It is important to observe that, by contrast with Eq.~\eqref{ZBS}, the border strip
  $\langle k_1,\dots,k_r\rangle$ appearing in the latter equations corresponds to a partition
  $\bk=(k_1,\dots,k_r)$ of length $N$. Since
  \begin{align*}
    d_0(\bk)+d_1(\bk)&=S^{(m|n)}_{\langle k_1,\dots,k_r\rangle,0}(1^m,1^n)+
    S^{(m|n)}_{\langle k_1,\dots,k_r\rangle,1}(1^m,1^n)\\&=S^{(m|n)}_{\langle
      k_1,\dots,k_r\rangle}(1^m,1^n)
    =d_A(\bk)
  \end{align*}
  by Eqs.~\eqref{SS01}, \eqref{Schurql} and~\eqref{dak}, Eq.~\eqref{Zqalt} admits an obvious
  interpretation in terms of the ``branching'' of type $A_{N-1}$ border strips. To wit, each
  border strip $\langle k_1,\dots,k_r\rangle$ with $\bk=(k_1,\dots,k_r)\in\cP_N$, whose degeneracy
  and energy for the HS chain of $A_{N-1}$ type are respectively $d_A(\bk)$ and
  $\sum_{i=1}^{r-1}\cE_A(K_i)$, splits into two different ``branched'' border strips
  $\langle k_1,\dots,k_r|0\rangle$ and $\langle k_1,\dots,k_r|N\rangle$ with respective
  degeneracies $d_0(\bk)$ and $d_1(\bk)$, and energies $\sum_{i=1}^{r-1}\cE(K_i)$ and
  $\sum_{i=1}^{r-1}\cE(K_i)+\cE(N)$. Moreover, by Remark~\ref{rem.Scomb} the degeneracies
  $d_0(\bk)$ and $d_1(\bk)$ of each of these branched border strips are respectively equal to the
  number of supersymmetric Young tableaux of types $\cT_0$ and $\cT_1$. This description of the
  spectrum of the $\su(m|n)$ supersymmetric HS chain of $BC_N$ type is in fact closely connected
  to the analogous one for the Polychronakos chain of $BC_N$ type deduced in Ref.~\cite{BS20}, the
  main difference between both models being that in the latter each type $A_{N-1}$ motif in
  general gives rise to $N+1$ branches with different energies instead of just two.
\end{rem}
\section{Examples}\label{sec.exa}

In this section we shall provide a few concrete examples illustrating the motif-based description
of the spectrum of the open supersymmetric HS chain~\eqref{Hchain} developed in the last section,
spelled out in the two rules (R1)-(R2) above.

\subsection{$\su(1|2)$, $N=3$}

Let us start with a simple example with $N=3$ spins, one bosonic and two fermionic degrees of
freedom. To begin with, since $n$ is even we have $n_{\vep'}=n/2=1$ regardless of the value of
$\vep'$. Thus the spectrum is independent of $\vep'$, which is obviously a general feature of the
model when $n$ is even. On the other hand, as we shall see next, the spectrum is highly dependent
on $\vep$.

\subsubsection{$\vep=+1$} In this case $m_\vep=n_{\vep'}=1$ and hence $*=2$, so that $B=\{2\}$ and
$F=\{1,3\}$.

\begin{table}[t]
  \centering
  \small
  \begin{tabular}[c]{|l|c|c|c|c|}
    \hline
    Partition& Motif& Tableaux& Energy& Degeneracy\\
    \hline\hline
    (4)& (0,0,0)&\vphantom{%
         \ytableausetup{smalltableaux}\ytableaushort{1,2,2,2,2}\en \ytableaushort{2,2,2,2}}%
         \ytableausetup{smalltableaux}\ytableaushort{1,2,2,2}\en \ytableaushort{2,2,2,2}& 0&2\\
    \hline
    (3,1)& (0,0,1) &\vphantom{ytableausetup{smalltableaux}\ytableaushort{\none1,\none1,\none2,23}}%
           \ytableausetup{smalltableaux}\ytableaushort{\none1,\none2,23}\en
                  \ytableaushort{\none2,\none2,23}& $\cE(3)$& 2\\
    \hline
    (2,2)& (0,1,0)& \makecell{\vphantom{\ytableausetup{smalltableaux}\ytableaushort{\none1,\none1,12,2\none}}
           \ytableausetup{smalltableaux}\ytableaushort{\none1,12,2\none}\en
                  \ytableaushort{\none2,12,2\none}\en \ytableaushort{\none1,13,2\none}\\[-5pt]
    \vphantom{\ytableausetup{smalltableaux}\ytableaushort{\none1,\none1,12,2\none}}
    \ytableaushort{\none2,13,2\none}\en \ytableaushort{\none1,23,2\none}\en
           \ytableaushort{\none2,23,2\none}}& $\cE(2)$& 6\\
    \hline
    (2,1,1)& (0,1,1)& \vphantom{\ytableausetup{smalltableaux}\ytableaushort{\none\none1,\none\none1,233}}
             \ytableausetup{smalltableaux}\ytableaushort{\none\none1,233}\en
             \ytableaushort{\none\none2,233}& $\cE(2)+\cE(3)$& 2\\
    \hline
    (1,3)& (1,0,0)& \vphantom{\ytableausetup{smalltableaux}\ytableaushort{11,2\none,2\none,2\ one}}%
           \ytableausetup{smalltableaux}\ytableaushort{11,2\none,2\none}\en
           \ytableaushort{12,2\none,2\none}\en
           \ytableaushort{13,2\none,2\none}\en\ytableaushort{23,2\none,2\none}& $\cE(1)$& 4\\
    \hline
    (1,2,1)& (1,0,1)& \vphantom{\ytableausetup{smalltableaux}\ytableaushort{\none11,23\none,2\none\none}}%
             \ytableausetup{smalltableaux}\ytableaushort{\none11,23\none}\en
    \ytableaushort{\none12,23\none}\en\ytableaushort{\none13,23\none}\en
    \ytableaushort{\none23,23\none}& $\cE(1)+\cE(3)$& 4\\
    \hline
    (1,1,2)& (1,1,0)& \makecell{\vphantom{\ytableausetup{smalltableaux}
             \ytableaushort{111,2\none\none,2\none\none}}%
             \ytableausetup{smalltableaux}\ytableaushort{111,2\none\none}\en
             \ytableaushort{112,2\none\none}\en\ytableaushort{113,2\none\none}\\[-4pt]
    \vphantom{\ytableausetup{smalltableaux}
    \ytableaushort{111,2\none\none,2\none\none}}%
    \ytableaushort{123,2\none\none}\en \ytableaushort{133,2\none\none}\en
    \ytableaushort{233,2\none\none}}
             & $\cE(1)+\cE(2)$& 6\\
    \hline
    (1,1,1,1)& (1,1,1)& \vphantom{\ytableausetup{smalltableaux}\ytableaushort{2333,2\none\none\none}}
               \ytableausetup{smalltableaux}\ytableaushort{2333}
                       & $\cE(1)+\cE(2)+\cE(3)$& 1\\
    \hline
  \end{tabular}
  \medskip
  \caption{Allowed motifs for the $\su(1|2)$ supersymmetric chain of $BC_N$ type~\eqref{Hchain}
    for $N=3$ spins and $\vep=+1$, with their corresponding energies and degeneracies.}
  \label{tab.motifs1}
\end{table}

In Table~\ref{tab.motifs1} we list all the partitions~$\bk=(k_1,\dots,k_r)$ of $N+1=4$, together
with their corresponding $(1|2)$ supersymmetric tableaux filled according to rules (R1)-(R2) in
the previous section (with $*=2$). Taking into account that for $N=3$
\[
  \cE(i)=i\bigg(3+\bar\be-\frac12(i+1)\bigg)=\frac i2\,(2\bar\be+5-i),
\]
we see that the chain's energies (in ascending order)
are in this case given by
\[
  0_2, (\bar\be+2)_4,\, (2\bar\be+3)_6,\, (3\bar\be+3)_2,\, (3\bar\be+5)_6,\,
  (4\bar\be+5)_4,\, (5\bar\be+6)_2,\, (6\bar\be+8)_1,
\]
where the subscripts indicate the corresponding degeneracies. In particular, since $\bar\be>0$ the
ground state (associated to the partition (4)) has zero energy and is twice degenerate. It easily
follows from the motif-based description of the spectrum that this last property is actually valid
for general $N$. Indeed, since $\cE(i)>0$ for $1\le i\le N$, it is clear from Eq.~\eqref{EdB} that
the ground state energy vanishes provided that the motif $\bde=(0^N)$, corresponding to the border
strip $\langle N+1\rangle$, is allowed. This is obviously the case when $m=1$, $n=2$, $\vep=1$,
since $*=m_\vep+n_{\vep'}=2\in B$ implies that the border strip $\langle N+1\rangle$ can be filled
according to rules (R1)-(R2) in the previous section by the two tableaux with bond vectors
$(2^{N+1})$ and $(1,2^{N})$. In particular, the ground state is twice degenerate in this case.
 
\subsubsection{$\vep=-1$} 
\begin{figure}[t]
  \centering
  \includegraphics[width=8.5cm]{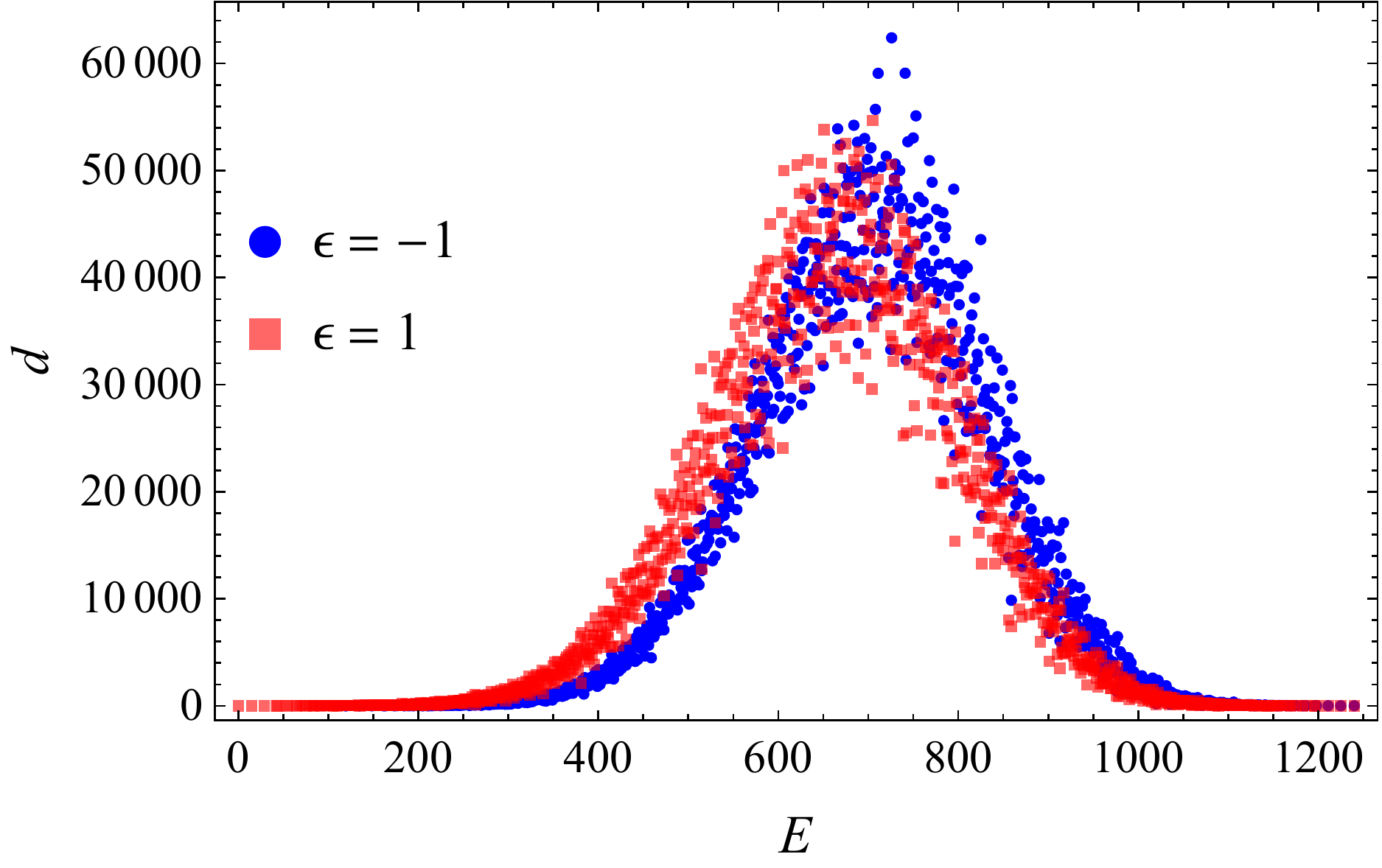}
  \caption{Energy $E$ vs.~degeneracy $d$ of the spectrum of the $\su(1|2)$ HS chain of $BC_N$ type
    with $N=15$ spins and $\bar\be=1$ for $\vep=-1$ (blue circles) and $\vep=1$ (pink squares).}
  \label{fig.Ed}
\end{figure}

Now $m_\vep=0$, $n_{\vep'}=1$ and thus $*=1$. This is exactly the
situation covered in Remarks~\ref{rem.mepone} and~\ref{rem.meptwo} in the previous section, since
necessarily $f_1=1$ but $*=1$ should be regarded as \emph{bosonic}. Thus a tableau like
$\ytableausetup{smalltableaux}\ytableaushort{11{\cdots},1\none\none}$ is \emph{allowed} in this
case, since the $1$ in the last (lowermost) box is regarded as bosonic in the comparison with the
one above it, while all the other $1$'s in the tableau are considered to be of fermionic type. For
the same reason, tableaux like $\ytableausetup{smalltableaux}\ytableaushort{11{\cdots}}$ are
forbidden. Taking (for instance) $F=\{1,2\}$ and $B=\{3\}$, and applying rules (R1)-(R2) above, it
is straightforward to show that the spectrum is given in this case (in order of ascending energy)
by
\[
  (2\bar\be+3)_4,\quad (3\bar\be+3)_4,\quad (3\bar\be+5)_5,\quad (4\bar\be+5)_8,\quad
  (5\bar\be+6)_4,\quad(6\bar\be+8)_2
\]
\begin{table}[t]
  \centering
  \small
  \begin{tabular}[c]{|l|c|c|c|c|}
    \hline
    Partition& Motif& Tableaux& Energy& Degeneracy\\
    \hline\hline
    (3,1)& (0,0,1)& \vphantom{ytableausetup{smalltableaux}\ytableaushort{\none1,\none1,\none2,13}}%
                    \ytableausetup{smalltableaux}\ytableaushort{\none1,\none2,13}\en
                    \ytableaushort{\none1,\none3,13}\en
                    \ytableaushort{\none2,\none3,13}\en
                    \ytableaushort{\none3,\none3,13}& $\cE(3)$& 4\en
    \\
    \hline
    (2,2)& (0,1,0)&\vphantom{\ytableausetup{smalltableaux}\ytableaushort{\none1,\none1,12,2\none}}
                    \ytableausetup{smalltableaux}\ytableaushort{\none1,12,1\none}\en
                    \ytableaushort{\none1,13,1\none}\en \ytableaushort{\none2,13,1\none}\en
                    \ytableaushort{\none3,13,1\none}& $\cE(2)$& 4\\
    \hline
    (2,1,1)& (0,1,1)& \vphantom{\ytableausetup{smalltableaux}\ytableaushort{\none\none1,\none\none1,233}}
                      \ytableausetup{smalltableaux}\ytableaushort{\none\none1,122}\en
                      \ytableaushort{\none\none1,123}\en
                      \ytableaushort{\none\none2,123}\en
                      \ytableaushort{\none\none3,123}& $\cE(2)+\cE(3)$& 4\\
    \hline
    (1,2,1)& (1,0,1)& \makecell{\vphantom{\ytableausetup{smalltableaux}%
                      \ytableaushort{\none11,23\none,2\none\none}}%
                      \ytableausetup{smalltableaux}\ytableaushort{\none11,12\none}\en
                      \ytableaushort{\none12,12\none}\en\ytableaushort{\none13,12\none}\en
                      \ytableaushort{\none11,13\none}\\[-4pt]
    \vphantom{\ytableausetup{smalltableaux}%
    \ytableaushort{\none12,13\none,2\none\none}}%
    \ytableausetup{smalltableaux}\ytableaushort{\none12,13\none}\en
    \ytableaushort{\none13,13\none}\en\ytableaushort{\none22,13\none}\en
    \ytableaushort{\none23,13\none}
    }& $\cE(1)+\cE(3)$& 8\\
    \hline
    (1,1,2)& (1,1,0)&\makecell{\vphantom{\ytableausetup{smalltableaux}
                      \ytableaushort{111,2\none\none,2\none\none}}%
                      \ytableausetup{smalltableaux}\ytableaushort{111,1\none\none}\en
                      \ytableaushort{112,1\none\none}\en\ytableaushort{113,1\none\none}\\[-4pt]
    \vphantom{\ytableausetup{smalltableaux}
    \ytableaushort{122,1\none\none,2\none\none}}%
    \ytableausetup{smalltableaux}\ytableaushort{122,1\none\none}\en
    \ytableaushort{123,1\none\none}\hfill\hfill
    }& $\cE(1)+\cE(2)$& 5\\
    \hline
    (1,1,1,1)& (1,1,1)& \vphantom{\ytableausetup{smalltableaux}\ytableaushort{2333,2\none\none\none}}
                        \ytableausetup{smalltableaux}\ytableaushort{1222}\en
                        \ytableaushort{1223}
                              & $\cE(1)+\cE(2)+\cE(3)$& 2\\
    \hline
  \end{tabular}
  \medskip
  \caption{Allowed motifs for the $\su(1|2)$ supersymmetric chain of $BC_N$ type~\eqref{Hchain}
    for $N=3$ spins and $\vep=-1$, with their corresponding energies and degeneracies. (We have
    taken~$F=\{1,2\}$ and $B=\{3\}$.)}
  \label{tab.motifs2}
\end{table}%
(cf.~Table~\ref{tab.motifs2}). We see that the degeneracies differ significantly from those in the
case~$\vep=+1$ listed in Table~\ref{tab.motifs1}. In particular, the two levels corresponding to
the partitions $(4)$ and $(1,3)$ are absent in this case. Moreover, the ground state is now
associated to the partition (2,2), has energy $2\bar\be+3\ge 3$ and is four times degenerate. For
arbitrary $N$, an analysis similar to the one above shows that the ground state corresponds to the
motif $(0^{N-3},1,0)$, or equivalently to the border strip $\langle N-1,2\rangle$, and has energy
$\cE(N-1)=(N-1)(\bar\be+N/2)$. The ground state is again four times degenerate, corresponding to
the four tableaux
\[
  \ytableaushort{\none1,\none2,\none{\raise -1pt\hbox{$\scriptscriptstyle\vdots$}},
    \none2,12,1\none}\qquad \ytableaushort{\none2,\none2,\none{\raise
      -1pt\hbox{$\scriptscriptstyle\vdots$}},\none2,12,1\none}\qquad
  \ytableaushort{\none1,\none2,\none{\raise
      -1pt\hbox{$\scriptscriptstyle\vdots$}},\none2,13,1\none}\qquad
  \ytableaushort{\none2,\none2,\none{\raise
      -1pt\hbox{$\scriptscriptstyle\vdots$}},\none2,13,1\none}
\]
allowed for the border strip $\langle N-1,2\rangle$ according to the rules (R1)-(R2) in the
previous section. Incidentally, for both $\vep=1$ and $\vep=-1$ the highest excited state, with
energy
\[
  \sum_{i=1}^N\cE(i)=\frac16\,N(N+1)(3\bar\be+2N-2),
\]
is obviously associated to the single-line border strip $\langle 1^{N+1}\rangle$, and is
nondegenerate for $\vep=1$ (the only allowed tableau is $\ytableaushort{23\cdots3}\,$) and twice
degenerate for $\vep=-1$ (the two allowed bond strips being $\ytableaushort{12\cdots2}$ and
$\ytableaushort{12\cdots23}\,$). In particular, the spectrum is less spread for $\vep=-1$ than for
$\vep=1$, as is already apparent from Fig.~\ref{fig.Ed} for the case $N=15$.

\medskip The description of the spectrum developed in the previous section makes it feasible to
exactly compute the spectrum of the $\su(m|n)$ HS chain of $BC_N$ type~\eqref{Hchain} for a
relatively large number of spins using standard symbolic packages. For instance, in
Fig.~\ref{fig.Ed} we present the result of the computation with \emph{Mathematica}\texttrademark\
of the spectrum of the $\su(1|2)$ chain with $\bar\be=1$ and $N=15$ spins for both $\vep=1$ and
$\vep=-1$ (recall that in this case the spectrum is independent of $\vep'$). Much as in the
$A_{N-1}$ case, both spectra show a very high degeneracy\footnote{In fact, since by
  Eqs.~\eqref{disprel} and~\eqref{EdB} the energies are of the form $i\bar\be+j$ with $i,j$
  nonnegative integers, it is clear that the degeneracy is higher when $\bar\be$ is a positive
  integer or rational number with a small denominator.} (of the order of $50\,000$ for energies
near the median) and a Gaussian-like shape. In particular, the high degeneracy of the spectrum and
the existence of a motif-based description thereof strongly suggest that this model possesses
twisted Yangian symmetry. As mentioned in the Introduction, the existence of this symmetry was
established in Ref.~\cite{BPS95} only in the non-supersymmetric case and for the three uniform
cases listed in Table~\ref{tab.sites}.

\subsection{$\su(1|1)$, arbitrary $N$}

The partition function of the $\su(1|1)$ HS chain of $A_{N-1}$ type can be computed in closed form
for arbitrary $N$, with the result~\cite{CFGRT16}
\[
  Z_{A,N}(q)=2\prod_{i=1}^{N-1}(1+q^{\cE_A(i)})\,,
\]
where we have explicitly indicated the dependence on $N$ for later convenience. A similar formula
is in fact valid for the $A_{N-1}$ Polychronakos--Frahm~\cite{Fr93,Po94} (rational) and
Frahm--Inozemtsev~\cite{FI94} (hyperbolic) chains, with $\cE_A$ replaced by the dispersion
relation of the latter chains. As in the $A_{N-1}$ case, the partition function of the $\su(1|1)$
HS chain of $BC_N$ type can be evaluated in closed form for arbitrary $N$, as we shall next show.
In particular, we shall see that the result depends in an essential way on the two signs $\vep$
and $\vep'$.

\subsubsection{$\vep=\vep'=1$}

In this case $m_\vep=n_{\vep'}=1$, and therefore $*=2$, $F=\{1\}$ and $B=\{2\}$. Since $2$ is
bosonic, no type 1 tableaux of the form
\[
  \ytableausetup{nosmalltableaux}
  \overset{^{\hphantom{}\ytableaushort{\none\none}}{}^{\textstyle\iddots}}%
  {\ytableaushort{\none{},\none{\raise-2pt\hbox{$\vdots$}},2{}}}
\]
are allowed. Thus all allowed tableaux are of type 0, i.e., of the form
\[
  \overset{^{\hphantom{}\ytableaushort{\none}}{}^{\textstyle\iddots}}%
  {\ytableaushort{{},{\raise-2pt\hbox{$\vdots$}},2}}
\]
It is also clear that the $2$ in the last box imposes no additional restriction (apart from the
standard rules for $(1|1)$ supersymmetric tableaux) on the box immediately on top of it. In other
words, the number of tableaux of this form with $N+1$ boxes coincides with the number of regular
$\su(1|1)$ tableaux with $N$ boxes obtained by removing the last (bottommost) box. We thus arrive
at the formula
\[
  Z_N^{++}(q)=\sum_{\bk\in\cP_N}d_A(\bk)q^{\sum_{i=1}^{r-1}\cE(K_i)}\,.
\]
Realizing that the RHS is nothing but $Z_{A,N}(q)$ with $\cE_A$ replaced by $\cE$ we conclude that
\[
  Z_N^{++}(q)=2\prod_{i=1}^{N-1}(1+q^{\cE(i)})\,.
\]
Thus the $\su(1|1)$ HS chain of $BC_N$ type with $\vep=\vep'=1$ behaves essentially as a type
$A_{N-1}$ chain with a different dispersion relation.

\subsubsection{$\vep=\vep'=-1$}

We have $m_\vep=n_{\vep'}=0$, and therefore $*=m_\vep+n_{\vep'}=0$. Thus only type 1 tableaux are
allowed, and it is again clear that the $0$ in the leftmost box entails no restriction (apart from
the standard rules for $(1|1)$ supersymmetric tableaux) on the box to its right. We thus have
\begin{align*}
  Z_N^{--}(q)&=\sum_{\bk\in\cP_N}d_A(\bk)q^{\cE(N)}q^{\sum_{i=1}^{r-1}\cE(K_i)}=q^{\cE(N)}Z^{++}_{N}(q)\\
             &=2q^{\cE(N)}\prod_{i=1}^{N-1}(1+q^{\cE(i)})\,.
\end{align*}
Thus the spectrum in this case is obtained by shifting the spectrum in the previous case by a
constant (positive) energy $\cE(N)$.

\subsubsection{$\vep=-\vep'=1$}

In this case $m_\vep=1$, $n_{\vep'}=0$, and hence $*=m_\vep+n_{\vep'}=1$, $B=\{1\}$ and $F=\{2\}$.
A moment's reflection shows that all border strips $\langle k_1,\dots,k_r\rangle$ give rise to
exactly one allowed Young tableau, of the form
\[
  \ytableausetup{smalltableaux}
  \overset{^{\hphantom{}\ytableaushort{\none\none\none\none\none\none}}{}^{\textstyle\iddots}}%
  {\ytableaushort{\none\none\none12,\none\none\none{\raise-1pt\hbox{$\scriptscriptstyle\vdots$}},
      \none\none\none1,12{\cdots}2,{\raise-1pt\hbox{$\scriptscriptstyle\vdots$}}\none,1\none,1\none}}\qquad
  \overset{^{\hphantom{}\ytableaushort{\none\none\none\none\none\none\none}}{}^{\textstyle\iddots}}%
  {\ytableaushort{\none\none\none\none\none\none1,\none\none\none12\cdots2,\none\none\none{\raise-1pt\hbox{$\scriptscriptstyle\vdots$}},
      \none\none\none1,12\cdots2}}
\]
respectively for type 0 and 1 border strips. Thus the partition function is given in this case by
\begin{align*}
  Z_N^{+-}(q)&=\sum_{\bk\in\cP_{N+1}}q^{\sum_{i=1}^{r-1}\cE(K_i)}=1+\sum_{r=2}^N\,\sum_{1\le
               K_1<\cdots<K_{r-1}\le N}q^{\sum_{i=1}^{r-1}\cE(K_i)}\\
  &=\prod_{i=1}^N(1+q^{\cE(i)}).
\end{align*}
Thus the $\su(1|1)$ HS chain of $BC_N$ type with $\vep=-\vep'=1$ is equivalent to a system of $N$
free spinless fermions with dispersion relation~$\cE$ (i.e., for which the energy of the
single-particle mode with momentum $2k\pi/N$ is $\cE(k)$).

\subsubsection{$\vep=-\vep'=-1$}

Here $m_\vep=0$, $n_{\vep'}=1$, and consequently $*=m_\vep+n_{\vep'}=1$, $F=\{1\}$ and $B=\{2\}$.
The difference with the previous case is that, even if now $1$ is of fermionic type, $*=1$ should
be treated as a bosonic variable (cf.~Remarks~\ref{rem.mepone}) and~\ref{rem.meptwo} above). As a
consequence, type 0 and type 1 tableaux can only end in $\ytableaushort{1\cdots,1\none}$ and
$\ytableaushort{\none\cdots,12}$, respectively, where in both cases $\ytableaushort{\cdots}$
stands for a standard $(1|1)$ supersymmetric tableaux with no additional restrictions. Since
$K_{r-1}=N-1+\al$ for type $\al$ tableaux, we conclude that
\[
  Z_N^{-+}(q)=(q^{N-1}+q^N)Z_{A,N-1}(q)\big|_{\cE_A\to\cE}=2\big(q^{\cE(N-1)}+q^{\cE(N)}\big)
  \prod_{i=1}^{N-2}(1+q^{\cE(i)}).
\]

\subsubsection{Free energy}

With the previous explicit formulas it is an easy matter to obtain an exact expression for the
free energy per spin of the open $\su(1|1)$ HS chain~\eqref{Hchain} in the thermodynamic limit. To
this end, we first normalize the Hamiltonian dividing it by $1/N^2$, in order to obtain a finite
energy density in the thermodynamic limit. Since
\[
  \frac{\cE(i)}{N^2}=x_i\bigg(1+\frac{\bar\be}N-\frac{x_i}2-\frac1{2N}\bigg),\qquad
  x_i:=\frac iN\,\in(0,1],
\]
we have
\[
  \frac{\cE(i)}{N^2}\underset{N\to\infty}{\to}\frac
  x2\,(2\ga-x)=:\vp(x)\,,\qquad\ga:=1+\lim_{N\to\infty}\frac{\bar\be}N\ge1\,,
\]
where $x\in[0,1]$ is a continuous variable. The free energy per particle in the thermodynamic
limit is then given in all four cases by
\[
  f(T)=-T\lim_{N\to\infty}\frac1N\log
  Z_N^{\vep\vep'}\Bigl(\e^{-1/N^2T}\Bigr)=-T\int_0^1\log\bigl(1+\e^{-\vp(x)/T}\bigr)\diff x\,.
\]
Changing to the momentum variable $p=\pi x$ we obtain
\begin{equation}
  \label{f11}
  f(T)=-\frac T\pi\int_0^\pi\log\bigl(1+\e^{-\phi(p)/T}\bigr)\diff p\,,
\end{equation}
where the dispersion relation $\phi(p)$ is given by
\begin{equation}
  \label{disprelcont}
  \phi(p):=\vp\Bigl(\tfrac{|p|}{\pi}\Bigr)=\frac{|p|}{2\pi^2}(2\pi\ga-|p|)=\frac1{2\pi^2}\big[\pi^2\ga^2-(|p|-\pi\ga)^2\big]\,,\en\quad -\pi\le p\le\pi.
\end{equation}
\begin{figure}[t]
  \small
  \def\gam{4/3}
  \def\yend{5/3}
  \begin{tikzpicture}[xscale=2.2,yscale=1.5*1.15]
    \draw[->,thick] (-1.3,0)--(1.3,0) node[anchor=west] {$p$};
    \draw[->,thick] (0,0)--(0,1.2*\yend) node[anchor=south] {$\phi(p)$};
    \draw[dashed] (-1,\yend)--(1,\yend);
    \draw[dashed] (-1,0)--(-1,\yend);
    \draw[dashed] (,0)--(1,\yend);
    \draw (-1,-.05)--(-1,.05);
    \node at (-1,-.12) {$-\pi$};
    \draw (1,-.05)--(1,.05);
    \node at (1,-.12) {$\pi$};
    \node at (.27,\yend+.17) {$\ga-\frac12$};
    \draw[blue, thick, domain=-1:1] plot (\x, {abs(\x)*(2*\gam-abs(\x))});
    \draw[blue, thick, dashed, domain=-1.3:-1] plot (\x, {abs(\x+2)*(2*\gam-abs(\x+2))});
    \draw[blue, thick, dashed, domain=1:1.3] plot (\x, {abs(\x-2)*(2*\gam-abs(\x-2))});
  \end{tikzpicture}
  \hfill
  \def\gam{1}
  \def\yend{1}
  \begin{tikzpicture}[xscale=2.2,yscale=10/4*1.15]
    \draw[->,thick] (-1.3,0)--(1.3,0) node[anchor=west] {$p$};
    \draw[->,thick] (0,0)--(0,1.2*\yend) node[anchor=south] {$\phi(p)$};
    \draw[dashed] (-1,\yend)--(1,\yend);
    \draw[dashed] (-1,0)--(-1,\yend);
    \draw[dashed] (,0)--(1,\yend);
    \draw (-1,-.05*.6)--(-1,.05*.6);
    \node at (-1,-.12*.6) {$-\pi$};
    \draw (1,-.05*.6)--(1,.05*.6);
    \node at (1,-.12*.6) {$\pi$};
    \node at (.28*.6,\yend+.17*.6) {$1/2$};
    \draw[blue, thick, domain=-1:1] plot (\x, {abs(\x)*(2*\gam-abs(\x))});
    \draw[blue, thick, dashed, domain=-1.3:-1] plot (\x, {abs(\x)*(2*\gam-abs(\x))});
    \draw[blue, thick, dashed, domain=1:1.3] plot (\x, {abs(\x)*(2*\gam-abs(\x))});
  \end{tikzpicture}
  \caption{Dispersion relation $\phi(p)=|p|(2\pi\ga-|p|)/2\pi^2$ of the $\su(1|1)$ supersymmetric
    HS chain of $BC_N$ type~\eqref{Hchain} in the thermodynamic limit for $\ga>1$ (left) and
    $\ga=1$ (right).}
  \label{fig.disp}
\end{figure}
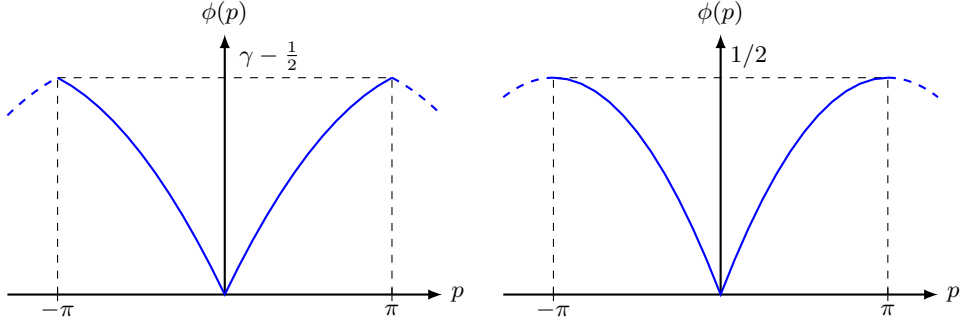%
Thus in the thermodynamic limit all four variants of the $\su(1|1)$ HS chain of $BC_N$ type are
equivalent to a system of free fermions with dispersion relation given by Eq.~\eqref{disprelcont}.
The latter expression is of course reminiscent of the corresponding ones for the free energy per
spin of the $A_{N-1}$-type HS, PF and FI chains obtained in Refs.~\cite{CFGRT16,FGLR18}. It is
clear that (when prolonged as a periodic function of period $2\pi$) $\phi(p)$ has a cusp at the
points $p=k\pi$ with $k\in\ZZ$ when $\ga>1$, or $2k\pi$ with $k\in\ZZ$ for $\ga=1$
(cf.~Fig.~\ref{fig.disp}), much like what happens with the dispersion relation of the
$A_{N-1}$-type PF and FI chains (when $\ga>1$) or the HS chain (when $\ga=1$). Since $\ga\ge1$,
the dispersion relation is clearly monotonic in each of the intervals $[-\pi,0]$ and $[0,\pi]$, as
is the case with the HS chains of $A_{N-1}$ type. Moreover, for $\ga=1$ Eq.~\eqref{disprelcont}
coincides (up to a trivial rescaling by a factor of $1/\pi^{2}$) with the dispersion relation of
the $\su(1|1)$ HS chain of $A_{N-1}$ type. This shows that the $\su(1|1)$ HS chain of $BC_N$ type
with $\bar\be/N\to0$ as $N\to\infty$ is equivalent in the thermodynamic limit to its $A_{N-1}$
counterpart, a result that is far from obvious a priori.

\section{Conclusions and outlook}\label{sec.conc}

The description of the spectrum of the Haldane--Shastry spin chain in terms of border strips (or,
equivalently, motifs) and skew Young tableaux is one of the hallmarks in the theory of integrable
spin chains with long-range interactions, underscoring the close connections of spin chains of HS
type with the representation theory of Yangian algebras. In this paper we address the problem of
finding a similar motif-based description of the spectrum of the open version of the
(supersymmetric) Haldane--Shastry spin chain, associated with the $BC_N$ root system. More
precisely, we first construct the model's Hamiltonian by suitably extending the standard
definition of the spin permutation and reversal operators to the supersymmetric case. We then
compute its partition function in closed form by means of Polychronakos's freezing trick, which
basically consists in modding out the dynamical degrees of freedom of the associated $BC_N$-type
spin Sutherland model. Inspired by the procedure for the closed ($A_{N-1}$) HS chain~\cite{BBH10},
we construct a generalized partition function depending polynomially on two sets of vector
variables, which reproduces the standard one when these variables are set equal to $1$. We then
show that this generalized partition function can be expressed in terms of two variants of the
standard skew super Schur polynomials, which can be defined through a simple combinatorial formula
in terms of supersymmetric skew Young tableaux with an additional box filled with a fixed integer.
With the help of this formula, we are able to derive a complete description of the spectrum of the
supersymmetric HS chain of $BC_N$ type in terms of extended motifs and restricted Young tableaux,
akin to the one for the closed HS chain. We illustrate this description with a few concrete
examples, including a complete study of the $\su(1|1)$ model and its thermodynamics.

Much as in the $A_{N-1}$ case, the existence of a motif-based description of the spectrum of the
model under study could prove of key importance for uncovering some of its fundamental properties.
In the first place, such a description is a clear indication that the model possesses some kind of
(twisted) Yangian symmetry. Obtaining an explicit realization of this symmetry, either via its
generators or through a suitable monodromy matrix~\cite{BPS95}, is certainly worth investigating.
As in the $A_{N-1}$ case~\cite{EFG12,FGLR18}, the motif-based description of the spectrum deduced
in this work can be taken as the starting point for deriving its thermodynamics using the
inhomogeneous transfer matrix approach. To this end, it is necessary to introduce a chemical
potential term in the Hamiltonian and generalize the above results ---in particular, the
characterization of the spectrum in terms of restricted supersymmetric skew Young tableaux--- to
the model thus obtained. In fact, the detailed results for the $\su(1|1)$ chains derived in this
paper strongly suggest that the thermodynamic functions in the general $\su(m|n)$ case can be
obtained from those of the closed supersymmetric HS chain simply by replacing the dispersion
relation of the latter model by that of the present one (cf.~Eq.~\eqref{disprel}). A related
application of our results is the study of the model's criticality by analyzing the low
temperature asymptotic behavior of its Helmholtz free energy, which should exhibit the $T^{2}$
behavior characteristic of $(1+1)$-dimensional conformal field theories~\cite{BCN86,Af86} at the
critical phase.

\section*{Acknowledgments}

This work was partially supported by Spain's Mi\-nis\-te\-rio de Ciencia, Innovaci\'on y
Universidades under grant PGC2018-094898-B-I00, as well as by Universidad Complutense de Madrid
under grant G/6400100/3000.

\address{Institut f\"ur Theoretische Physik, Universit\"at Innsbruck, Innrain 52, 6020 Innsbruck,
  Austria}

\noindent\email{Jose.Carrasco@uibk.ac.at} (JC)

\address{Departamento de F\'\i sica Te\'orica, Universidad Complutense de Madrid, Plaza de las
  Ciencias 1, 28040 Madrid, Spain}
  
\noindent\email{ffinkel@ucm.es {\normalfont (FF)}, artemio@ucm.es {\normalfont (AGL)},
  rodrigue@ucm.es} (MAR)

\end{document}